\documentclass[a4paper,fleqn]{cas-dc}

\usepackage[authoryear]{natbib}
\usepackage{algorithm}
\usepackage{algpseudocode}
\usepackage{graphicx}
\usepackage{ import } % Figure folder
\usepackage{ multirow }
\usepackage{ caption, subcaption }
\usepackage{listings} % Code snippet
\usepackage{xspace} % Matlab symbol
\usepackage{xpatch} % Code inline
\usepackage{xcolor}
\usepackage{todonotes}
\usepackage{realboxes}
\usepackage{cleveref}  % for clever reference
\definecolor{mygray}{rgb}{0.8,0.8,0.8}
\makeatletter
\xpretocmd\lstinline{\Colorbox{mygray}\bgroup\appto\lst@DeInit{\egroup}}{}{}
\makeatother

\SetSymbolFont{letters}{bold}{OML}{cmm}{b}{it}
\SetSymbolFont{operators}{bold}{OT1}{cmr}{bx}{n}

\newcommand{\norm}[1]{\lVert#1\rVert}   % Abs = norm
\newcommand{\lpare}[1]{\left(#1\right)} % Auto parentheses
\crefname{table}{Table}{Tables}% 
\crefname{equation}{Eq.}{Eq.}% 
\crefname{figure}{Fig.}{Fig.}%
\crefname{algorithm}{Algorithm}{Algorithm}%
\crefname{section}{Section}{Section}
\crefname{appendix}{}{}
\creflabelformat{section}{#2#1#3}
\crefname{subsection}{Section}{Section}
\creflabelformat{subsection}{#2#1#3}
\crefname{subsubsection}{Section}{Section}
\creflabelformat{subsubsection}{#2#1#3}

\newcommand{\dt}{\mathrm{d}t}

\newcommand{\xdot}{\dot{x}}
\newcommand{\lpp}{L_{\mathrm{pp}}}
\newcommand{\uvelo}{u_{\mathrm{s}}}
\newcommand{\uvelodot}{\dot{u}_{\mathrm{s}} }
\newcommand{\vm}{v_{\mathrm{m}}}
\newcommand{\vmdot}{\dot{v}_{\mathrm{m}}}

\newcommand{\xG}{x_{\mathrm{G}}}

\newcommand{\sumstate}{\sum_{i=1}^{6}}

\newcommand{\tfpla}{t_{\mathrm{f}}}
\newcommand{\tf}{\lpare{\tfpla}}

\newcommand{\tkpare}{\lpare{t_k}}

\newcommand{\cospsi}{\cos{\psi}}
\newcommand{\sinpsi}{\sin{\psi}}

\usepackage{mathtools}
\newcounter{mysubequations}

\usepackage{ulem}

\newcommand\Erase{\bgroup\markoverwith{\textcolor{red}{\rule[.5ex]{2pt}{0.4pt}}}\ULon}
\def\tsc#1{\csdef{#1}{\textsc{\lowercase{#1}}\xspace}}
\tsc{WGM}
\tsc{QE}
\ExplSyntaxOn
\keys_set:nn{ stm / mktitle }{ nologo }
\ExplSyntaxOff

\begin{document}
\let\WriteBookmarks\relax
\def\floatpagepagefraction{1}
\def\textpagefraction{.001}

\shorttitle{Practical Berthing Trajectory Planner with Speed Control}    
\shortauthors{Agnes N. Mwange et~al.}  

\title [mode = title]{A Practical and Online Trajectory Planner for Autonomous Ships' Berthing, Incorporating Speed Control}  

%\tnotemark[<tnote number>] 

%\tnotetext[<tnote number>]{<tnote text>} 


% List of authors
\author[1,2]{Agnes N. Mwange}[]
\cormark[1]
\ead{mwange_agnes_ngina@naoe.eng.osaka-u.ac.jp}
\credit{Conceptualization, Methodology, Software, Writing - original draft}

\author[3]{Dimas M. Rachman}[]%[]
%\ead{}
\credit{Methodology, Software}

\author[1]{Rin Suyama}[]
\credit{Visualization, Writing - review \& editing}
\author[1]{Atsuo Maki}[]
\ead{maki@naoe.eng.osaka-u.ac.jp}
\credit{Supervision, Writing - review \& editing, Funding acquisition}

\address[1]{Department of Naval Architecture and Ocean Engineering, Osaka University, 2-1 Yamadaoka, Suita, 565-0871, Osaka, Japan}
\address[2]{Department of Marine Engineering and Maritime Operations, Jomo Kenyatta University of Agriculture and Technology (JKUAT), P.O. Box 62000 - 00200, Nairobi, Kenya}
\address[3]{Intelligent Processing Technologies Laboratory, Furuno Electronic Co., Ltd, 8-1 Jingikan-cho, Nishinomiya, 662-0843, Japan}
\cortext[cor1]{Corresponding author}

\begin{abstract}
Autonomous ships are essentially designed and equipped to perceive their internal and external environment and subsequently perform appropriate actions depending on the predetermined objective(s) without human intervention. Consequently, trajectory planning algorithms for autonomous berthing must consider factors such as system dynamics, ship actuators, environmental disturbances, and the safety of the ship, other ships, and port structures, among others.  In this study, basing the ship dynamics on the low-speed MMG model, trajectory planning for an autonomous ship is modeled as an optimal control problem (OCP) that is transcribed into a nonlinear programming problem (NLP) using the direct multiple shooting technique. To enhance berthing safety, besides considering wind disturbances, speed control, actuators' limitations, and collision avoidance features are incorporated as constraints in the NLP, which is then solved using the Sequential Quadratic Programming (SQP) algorithm in MATLAB. Finally, the performance of the proposed planner is evaluated through (i) comparison with solutions obtained using CMA-ES for two different model ships, (ii) trajectory planning for different harbor entry and berth approach scenarios, and (iii)  feasibility study using stochastically generated initial conditions and positions within the port boundaries. Simulation results indicate enhanced berthing safety as well as practical and computational feasibility making the planner suitable for real-time applications.
\end{abstract}

%Research highlights
%\item An optimal-control-based online trajectory planner for autonomous ships' berthing is proposed. 
%\item Speed reduction criteria and actuator limitation are incorporated into the planner.
%\item The algorithm is validated through comparison with solutions obtained using CMA-ES.
%\item Performance of the algorithm is demonstrated through trajectory planning for different harbor entry and berth approach scenarios, and feasibility study using stochastically generated initial conditions.

% Keywords
% Each keyword is separated by \sep

\maketitle
\sloppy

% Main text
\section{Overview} \label{Sec:1-overview}
During the berthing process, the ship is operated at a very low surge velocity, which renders the rudder ineffective in steering the ship through the desired intricate maneuvers. Moreover, besides reduced maneuverability, the complexity of the berthing process can also be attributed to the large drift angles, highly nonlinear dynamics, environmental disturbances, and complex harbor environments.
As demonstrated by \cite{ohtsu_takai_yoshihisa_1991}, the berthing process can be divided into two stages; (i) from harbor entry until the ship reaches the docking pose, and (ii) crabbing towards the final berthing point. The first stage involves speed reduction and a controlled approach to the docking pose, usually parallel to the berth and a few meters away from it, that may be achieved by the captain or through tugboat assistance. The second stage involves lateral movement from the docking pose to the final berthing position. This may be achieved using mooring lines, tugboats, or side thrusters. However, when done manually, these practices are time-consuming, tedious, and prone to miscommunication which results in devastating accidents and losses. On the other hand, an autonomous ship must successfully and efficiently achieve these intricate maneuvers with minimal human assistance. Subsequently, to achieve autonomous berthing, it is essential to design intelligent control systems that integrate vital technologies and algorithms such as sensors, navigation algorithms, trajectory planning, collision avoidance strategies, and closed-loop control, which orchestrate the vessel's movements with precision and safety. This study focuses on trajectory planning because of its crucial role in determining the optimal route and ensuring safe berthing.

\subsection{Related Research}
Trajectory planning for autonomous berthing can be considered a  constrained nonlinear optimization problem that necessitates determining the optimal inputs that will steer the ship from an arbitrary position to the desired berthing position while considering ship dynamics, desired objective(s), path constraint(s), and environmental disturbances where possible. Solving this optimization problem to generate dynamically and practically feasible states and control trajectories is a testament to successful trajectory planning. Unfortunately, a review done by \cite{OZTURK2022111010} shows that only about $ 1\% $ of the existing trajectory planning algorithms focus on autonomous berthing. This indicates the dire need for further and extensive research on autonomous berthing to reinforce the concept's viability in practical applications.

Path and trajectory planning algorithms can be classified into two major categories as presented by \cite{sym10100450}, (i) local optimization, and (ii) global optimization algorithms. The two have distinguished scopes and approaches and their application is dependent on the characteristics and specific requirements of the optimization problem. In trajectory planning for autonomous ships, these methods have been used individually or in a combined form to achieve outstanding results. From among the global optimization algorithms is the path finding A$^*$ algorithm proposed by \cite{hart1968formal}, \cite{hart1972correction}. The algorithm has been used in modified forms as presented by \cite{article} with a focus on collision avoidance and \cite{YUAN2023113964} with a focus on improved path smoothness and trackability. Although fast and efficient, its performance is highly dependent on the quality of the heuristic function. Another approach to global trajectory optimization is customizing powerful optimization algorithms to trajectory optimization. \cite{maki2020application} proposed a berthing trajectory planning algorithm based on the Covariance Matrix Adaptation Evolution Strategy, (CMA-ES) proposed by \cite{inbook}, and tailored to handle inequality constraints as presented by \cite{sakamoto2017modified}.  Further improvements and modifications to realize practically realistic control input were presented by \cite{maki2021application}.  Moreover, CMA-ES was employed with redefined objectives to ensure safe trajectories in realistic ports presented by \cite{MIYAUCHI2022110390}  and to replicate the captain's maneuver at port presented by \cite{SUYAMA2022112763}. Although its high computation cost limits it to offline applications, the evolutionary algorithm is robust and reliable for trajectory optimization applications. 

Local optimization algorithms are among the earliest works in automating the berthing process. \cite{koyam1987systematic} proposed a knowledge-based planner and controller capable of handling environmental disturbances. Later on, \cite{yamato1993automatic} proposed an expert system planner rules, and constraints governing the decision-making system were based on empirical data and information.  Further on, using indirect numerical methods, \cite{shouji1992automatic} formulated the berthing problem as a nonlinear two-point boundary value problem (TPBVP) and solved it using the sequential conjugate gradient restoration (SCGRA) algorithm proposed by \cite{wu1980sequential}. Due to the high computation cost, the algorithm is limited to offline applications. \cite{380251} modeled the berthing problem as a minimum time and energy nonlinear programming problem using the discrete Augmented Lagrangian approach. The main objective of this approach was to improve the control input as well as substantial consideration of system dynamics, non-linearities, and environmental disturbances in trajectory planning. Besides the computation requirements limiting it to offline application, the scope of this work is limited to the first stage of the berthing process. The Artificial Potential Field (APF) algorithm proposed by  \cite{khatib1986real} has been exploited in modified forms as presented by \cite{LIAO201947} for enhanced smoothness in heading and \cite{HAN2022112877} for collision avoidance with dynamic obstacles. These works highlight the potential of pairing APF with other algorithms in enhancing (Convention on the International Regulations for Preventing Collisions at Sea) COLREGs compliance and real-time performance. \cite{3d3470386562d531c40e187b8160d335d1012412} incorporated multiple shooting technique into Nonlinear Model Predictive Control (NMPC) and successfully designed an auto-berthing controller. Further, based on NMPC \cite{ZHANG2023115228} proposed a path planning algorithm for minimal berthing time. Although the algorithm can guarantee minimal berthing time, collision avoidance and environmental disturbances are not considered. \cite{martinsen2020optimization} proposed a collocation-based planner whose performance for real-time application was reinforced by implementing an NMPC-based tracking controller.

Moreover, in a combined form, potential solutions from global optimization algorithms have been used to warm start local optimization algorithms thereby enhancing accuracy and computation efficiency. In most of these cases, the trajectory optimization problem is considered as an OCP, that is transcribed into an NLP using suitable discretization schemes and solved appropriately. \cite{BITAR2019308} used direct collocation to refine and improve the accuracy of a potential solution obtained using the A$^*$ algorithm. With a focus on computation efficiency, \cite{rachman2022warm} transcribed the OCP into an NLP using direct collocation and a CMA-ES solution as the warm start.  With a focus on the quality of the warm start solution, \cite{WANG2023114156} transcribed the OCP into an NLP using direct multiple shooting and used a solution from an artificially guided Hybrid A$^*$ algorithm as the warm start.

It is noteworthy that previous research on berthing trajectory planning has largely overlooked the navigational safety perspective, with safety often being treated as a secondary concern as highlighted by \cite{OZTURK2022111010}. This has significantly affected the rate of implementation of autonomous berthing. Therefore, integrating safety considerations more proactively into trajectory planning methods could play a crucial role in the future of autonomous berthing and its potential to coexist with conventional berthing methods.

\subsection{Proposed Planner}
The proposed planner is based on local optimization with a focus on enhancing berthing safety and accuracy while ensuring reasonable computation speed and independence on the quality of the initial guess. The trajectory optimization problem is modeled as a minimum-time optimal control problem (OCP), transcribed into a nonlinear programming problem (NLP) using direct multiple shooting, and solved using the fmincon solver, Sequential Quadratic Programming (SQP) algorithm in MATLAB. The contributions of this study are :
\begin{enumerate}
    \item An online optimal-control-based planner that uses a simple linear guess (linear interpolation between initial and terminal optimization variables) to initialize the SQP algorithm to obtain feasible solutions.
    \item Practical consideration of actuators in trajectory planning through imposing an artificial limit on the actuators as well as incorporating a rate of change of the actuators based on actual data.
    \item Incorporation of speed reduction guidelines in autonomous berthing for enhanced berthing safety.
    \item Reasonable computational cost making the planner suitable for re-planning and real-time applications.
\end{enumerate}

To validate the algorithm, the optimal states' trajectories obtained using the proposed planner are compared to those obtained using CMA-ES for two different model ships: ship A and ship B. Ship A is the subject model ship, but with artificially limited actuation, while Ship B is an underactuated model ship, as detailed in \cref{tab: model ship particulars}.

The rest of the paper is organized as follows: Section 2 introduces the important notations used in the paper; Section 3 introduces the model ship and its mathematical model. This section further introduces the OCP to be solved, the transcription of the OCP to an NLP, and the detailed description of the NLP constraints. Transcription, actuator, speed reduction, and collision avoidance constraints are described here.  The section also shows the simulation conditions used to validate and demonstrate the performance of the algorithm. The simulation results are shown in Section 4 and discussed in Section 5; Section 6 concludes the study.

\section{Notations}\label{sec:notations}
In this study, $\mathbb{R}^n$ denotes the $n$-dimensional Euclidean space where every coordinate is a real number. The $\odot$ symbol denotes the Hadamard product, that is, the element-wise multiplication of vectors of the same size where the output vector is of the same size as the input vectors.

\section{Method}
\subsection{Model Ship}\label{sec: model ship}
\begin{figure}[htbp]
    \centering
     \includegraphics[width =\columnwidth]{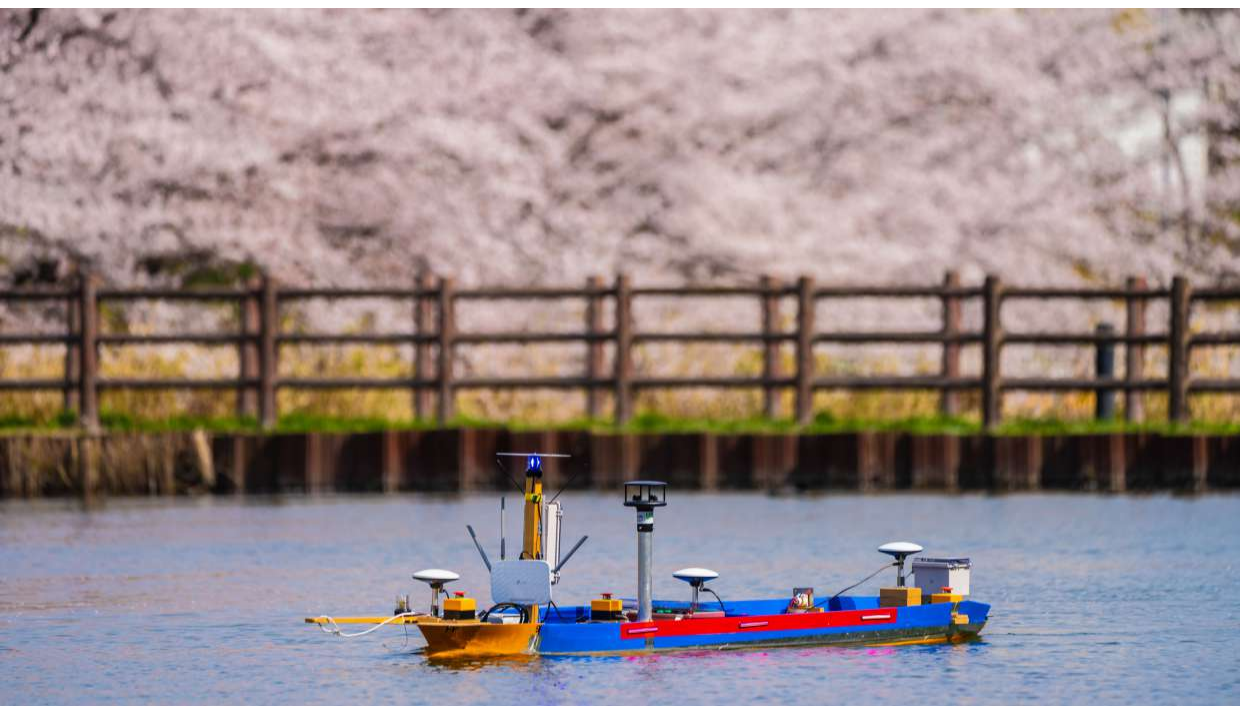}
    \caption{ Model Ship A at Inukai Pond, Osaka University }
    \label{fig: takaoki maru}
\end{figure}

\begin{table}[htbp] 
\caption{Model Ships' Principal Particulars} 
\label{tab: model ship particulars}
     \begin{tabular}{p{22mm}|p{22mm} |p{22mm}}
     \hline
     & Ship A & Ship B\\
     \hline \hline
        %\multicolumn{4}{ c }{Principal Dimensions}\\
       % \hline
        Length, $L$&  3.0m  & 3.0m \\
       
        Breadth, $B$ & 0.4m  & 0.4m \\
      
        Draft, $d$ & 0.17m  & 0.17m  \\ 
       
        Propeller & 1 fixed pitch propeller & 1 fixed pitch propeller\\
      
        Rudder & Vectwin rudder system & Single rudder\\
    
        Side Thrusters & 1 fixed pitch bow thruster & \\[ 1ex]
    \hline 
    \end{tabular}
 \end{table}

\subsection{ Maneuvering Model of the Ship}
\begin{figure}[ht]
    \centering
    \includegraphics[width=\columnwidth]{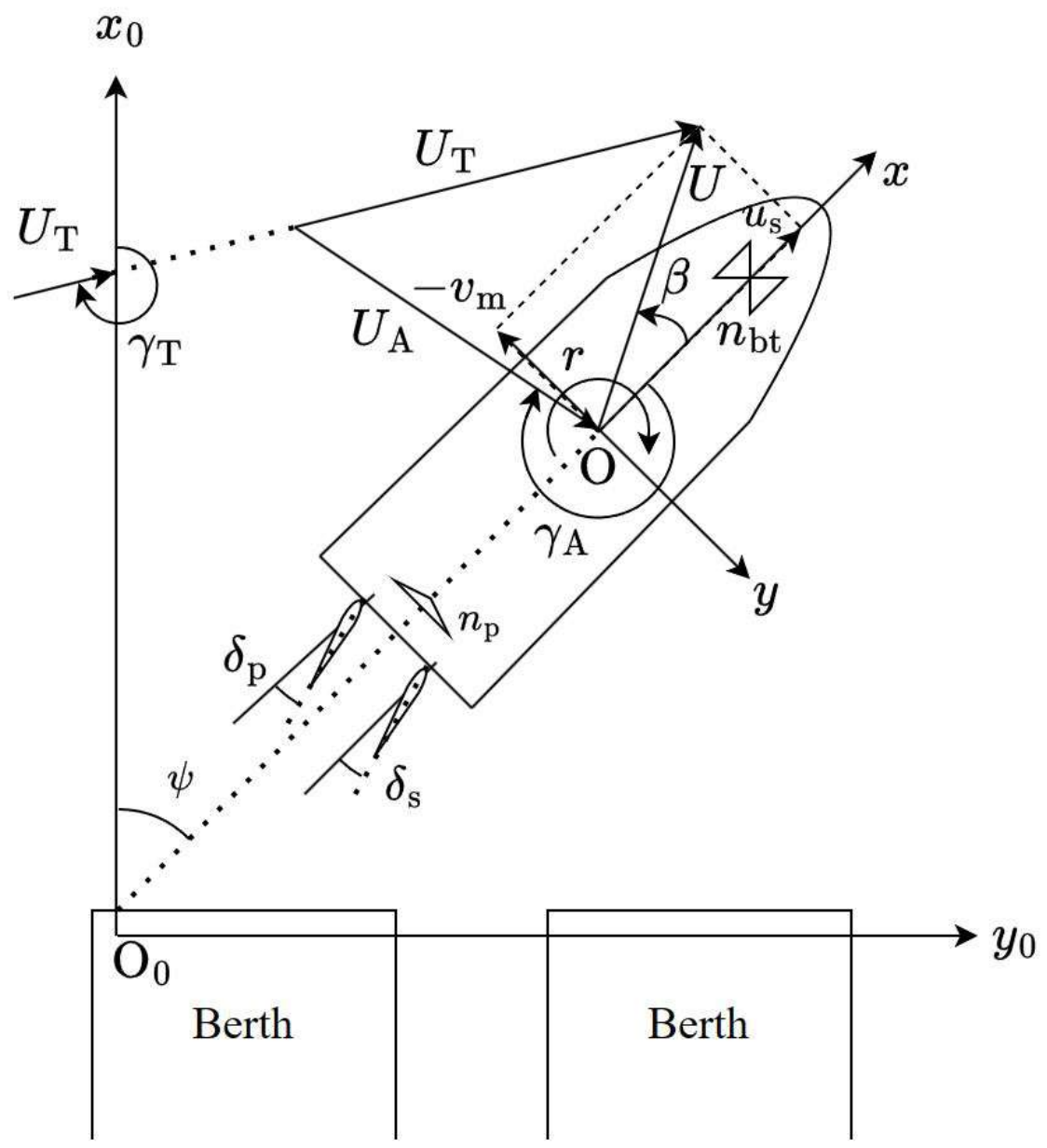}
    \caption{Coordinate systems}
    \label{fig: coordinate systems}
\end{figure}

Within the harbor areas, the ship is operated at a considerably low forward speed. Near the berth, in the presence of tugboats and side thrusters or not, besides large drift angles, the sway and yaw velocities are higher than the surge velocity. Consequently, it becomes extremely difficult to predict the components of maneuvering forces.  Accordingly, in this study, to compute the forces acting on the hull, the maneuvering model group (MMG) harbor maneuvering model proposed by \cite{Yasukawa2015} is used.
The kinematic ship model is a 3-DOF model that uses two coordinate systems: the earth-fixed system \(\mathrm{O}_0 - x_0, y_0, z_0\) and the ship-fixed system \(\mathrm{O} -x, y, z\) where \(\mathrm{O}\) is the ship's center of gravity as shown in \cref{fig: coordinate systems}. 
$ U_{\mathrm{T}} $ is the speed of the true wind and $ \gamma_{\mathrm{T}}$ is the direction of the true wind, positive in the clockwise direction starting from the x-axis of the earth-fixed coordinate system. $\beta $ is the ship's drift angle measured in the ship-fixed coordinate system.
$\psi, \uvelo, \vm$, and $ r$ denote the ship's yaw angle, surge velocity, sway velocity, and yaw velocity, respectively.  $U = \sqrt{ \uvelo ^ {2} + \vm ^ {2}}$, is the resultant ship speed. 

Further, \cref{eq: earth-ship coord} describes how the kinematics at the midship can be transformed from the ship-fixed coordinate system to the earth-fixed coordinate system.
\begin{equation}
  \begin{bmatrix} \dot{x}_0 \\ \dot{y}_0 \\ \dot{\psi} \end{bmatrix} =
  \begin{bmatrix} \cospsi & -\sinpsi &  0 \\ \sinpsi & \cospsi & 0 \\  0 & 0 & 1 \end{bmatrix} \begin{bmatrix} \uvelo\\ \vm \\ r  \end{bmatrix}  
  \label{eq: earth-ship coord}
\end{equation}

The equations of motion, as shown in \cref{eq: ma = F}, are based on the MMG model proposed by \cite{Yasukawa2015}.
\begin{equation}  
    \begin{split}
       (m + m_x)\uvelodot - (m + m_y) \vm r - \xG m r ^ 2 &= X   \\       
       (m + m_y)\vmdot+ (m + m_x) \uvelo r + \xG m \dot{r} &= Y \\        
       \lpare{{I_z}_z + {J_z}_z + \xG m ^ 2}\dot{r} + \lpare{\vm + \uvelo r}\xG m &= N_\mathrm{m},  \\         
    \end{split}
    \label{eq: ma = F}
\end{equation}

where $X$ and $Y$ denote the total surge and sway forces, and $N_\text{m}$ is the yaw moment about the midship. The right-hand-side of \cref{eq: ma = F} can further be expanded as follows:
\begin{equation} \label{eq: expanded forces and moment}
    \begin{split}
        X &= X_\mathrm{H} + X_\mathrm{P} + X_\mathrm{BT} + X_\mathrm{R} + X_\mathrm{A}\\
        Y &= Y_\mathrm{H} + Y_\mathrm{P} + Y_\mathrm{BT}+ Y_\mathrm{R}  + Y_\mathrm{A}\\
        N_\mathrm{m} &= N_\mathrm{H} + N_\mathrm{P} + N_\mathrm{BT} + N_\mathrm{R}  + N_\mathrm{A}\\      
    \end{split}
\end{equation}
The subscripts H, P, BT, R, and A denote hull, propeller, bow thruster, rudder, and air, respectively. \cref{eq: expanded forces and moment} summarizes the summation of surge and sway forces and moments resulting from hydrodynamic forces acting on the hull, steering forces and moments induced by the rudder, thrust forces and moments generated by the propeller and bow thruster, and wind forces and moments. The hydrodynamic forces and moments acting on the hull as well as forces and moments induced by the propeller thrust are expressed based on \cite{yoshimura2009unified} model. However, unlike conventional ships where the propeller can operate in both forward and reverse directions, propellers for ships equipped with a vectwin rudder system are operated in the forward direction only. The vectwin rudder system includes a pair of rudders with specially designed sectional profiles attached symmetrically on the ship's hull behind the propeller. The angle range for each of the rudders is $140 ~ \mathrm{deg.}$, that is, $105 ~ \mathrm{deg.}$ towards the outboard and $35 ~ \mathrm{deg.}$ towards the inboard. This enables the ship to achieve common maneuvers such as turning, emergency stop, and crash-astern, by maintaining a constant forward propeller revolution and a desirable combination of rudder angles. The rudders can be operated as a pair or individually as in a conventional ship with two rudders. The forces and moments induced by the rudder are formulated as presented by \cite{kang2008mathematical}. The forces and moments as a result of wind (air) are calculated using the method presented by \cite{fujiwara1998estimation}.

From \cref{eq: ma = F}, the expressions for mass ($m$, and added masses ($m_x$ and $m_y$) can be simplified as, $m + m_x = M_x, m + m_y = M_y$. Additionally, the expression $ ({I_z}_z + {J_z}_z + \xG ^ 2 m)$ is simplified as $ I_{z\text{m}}$. \\ Subsequently, \cref{eq: ma = F} can be rewritten as:
\begin{equation}
    \begin{split}
        \uvelodot &= \frac{( X + M_y \vm r + x_\mathrm{G}m r ^ 2)}{M_x}\\
       \vmdot  &= \frac{ (Y - M_x \uvelo r)I_{z\text{m}} - (N_\mathrm{m} - \xG m \uvelo r)(\xG m)}{ M_xI_{z\text{m}} - (\xG m) ^ 2 }\\     
        \dot{r} &= \frac{ (Y - M_x \uvelo r)(\xG  m) - (N_\mathrm{m} - \xG m \uvelo r ) M_y } {(\xG m) ^ 2 - M_yI_{z\text{m}}  }\\
        \label{eq: ma = F second}
    \end{split}
\end{equation}

\subsection{Optimal Control Problem}
The system dynamics as shown in \cref{eq: dynamics} are obtained by combining \cref{eq: earth-ship coord} and \cref{eq: ma = F second}.
\begin{equation}
\xdot(t) = f (x(t), u(t), U_\mathrm{T}, \gamma_\mathrm{T}  ),
\label{eq: dynamics}
\end{equation}
 where $U_\mathrm{T}$ and $\gamma_\mathrm{T}$ are the true wind speed and direction respectively. For the trajectory planner, wind parameters are taken at the beginning and assumed constant throughout the trajectory. Let $\Bar{t}$ be the actual time outside the planner. Wind parameters are taken at time $\Bar{t}_j$ which corresponds to $t = 0$ of the planner.

The states vector is defined as: 
\begin{equation}
    x(t) \equiv {[ x_0, y_0, \psi, \uvelo, v_\mathrm{m}, r ]}^ \intercal \in \mathbb{R}^6, 
    \label{states}
\end{equation}
  where $x_0$ and $y_0$ denote the earth-fixed position in the $x$ and $y$ directions, respectively.
 
 The control vector is defined as: 
 \begin{equation}
     u(t) \equiv {[ \delta_{\mathrm{p}},\delta_{\mathrm{s}}, n_{\mathrm{p}}, n_{\mathrm{bt}} ] }^ \intercal \in \mathbb{R}^4, 
 \end{equation}
where $ \delta_{\mathrm{p}},\delta_{\mathrm{s}}, n_{\mathrm{p}}, n_{\mathrm{bt}} $ denotes the port rudder, starboard rudder, propeller, and bow-thruster, respectively. 

\subsubsection{Direct Multiple Shooting Technique}
The direct numerical methods for solving OCPs can be classified into three; direct shooting, direct multiple shooting, and collocation methods as presented by \cite{betts1998survey}. Direct multiple shooting as presented by \cite{BOCK19841603} performs better than single shooting when dealing with nonlinear systems and combines the advantages of the simultaneous and sequential direct methods as presented by \cite{diehl2006fast}. Further, basic comparative studies done by
\cite{PUCHAUD2023116162} and \cite{garcia2014comparison} corroborate that, in nonlinear optimization,  the local collocation method is preferable for faster computation, while global collocation and direct multiple shooting method are preferable if accuracy takes precedence over computation time. Henceforth, direct multiple shooting is used in this study.

To discretize the continuous-time optimal control problem and transcribe it into a nonlinear programming problem (NLP), the trajectory with time interval $[ 0, t_\mathrm{f}]$ is divided into $N_s$ segments. Let $ i $ denote the $ i_{\mathrm{th}} $ segment such that, $ i = 1, 2,\dots,N_s $. The endpoints of the segments act as the discretization points and are hereafter known as $knots$.  Let $N_k$ denote the number of knots such that $ k = 1, 2,\dots, N_k $ where $ k $ denotes the $k_{\mathrm{th}} $ knot point. Additionally, $ N_k = N_s + 1 $.

 \begin{figure}[htbp]
     \includegraphics[width=\columnwidth]{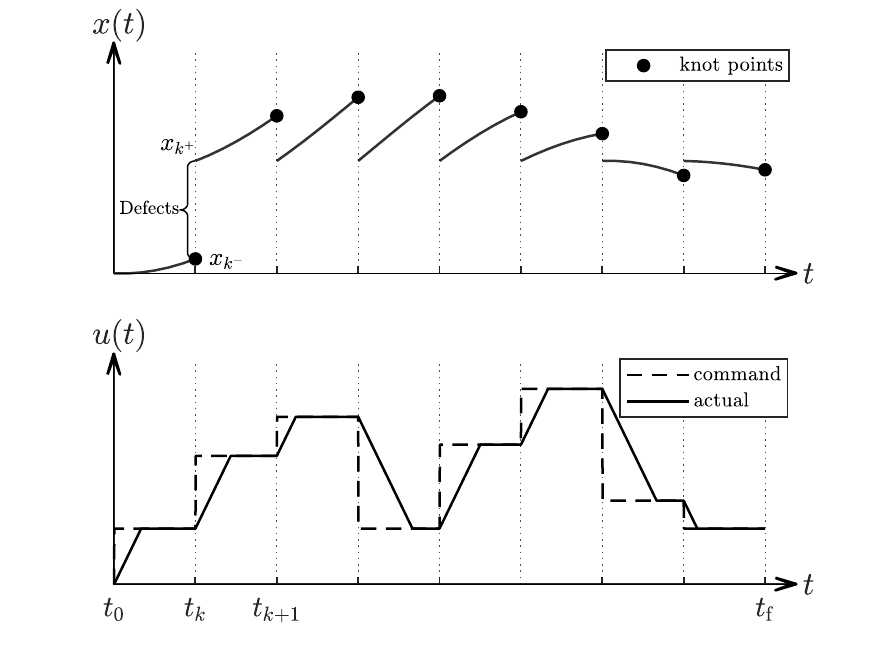}
    \caption{Direct Multiple Shooting Technique} \label{ fig: direct multiple shooting}
 \end{figure}  
The command control input per segment is approximated as piecewise constant, measured at knot points, $u_k \equiv u(t_k)$.  On the other hand, the actual control input used in the planner includes a rate of change and is saturated at the command input's value as shown in \cref{ fig: direct multiple shooting}. The rate of change is based on actual actuators' data. Although final stage actuation is supplied by tug boats, \cite{kose1986computer}  demonstrates the necessity of imposing actuator limits in the planning stage. Deliberately limiting the control input enhances berthing safety by creating a buffer zone that allows the system to counter unknown disturbances. Henceforth, in this study, only $43\%$ of the rudders', $50\%$ of the propeller, and $75\%$ of the bow thruster actuation are used.
For any segment $i$ starting from $k$ to $k+1$ knot points, the states $x_k \equiv x(t_k)$ and control $ u_k $ obtained from each iteration are used to integrate the systems equations of motion,  (true dynamics), $f_\mathrm{T}(t_k)$ over that segment using 4th order Runge-Kutta Scheme. 

\begin{equation}
  x_\mathrm{T}(t_{k+1}) =\int _{t_k}^{t_{k+1}}f_\mathrm{T}\lpare{x(t_k), u(t_k), U_\mathrm{T}(\Bar{t}_j), \gamma_\mathrm{T}(\Bar{t}_j) } \dt 
  \label{eq: True dynamics integration}
\end{equation}
 The 'true states', $x_\mathrm{T}(t_{k+1})$  obtained from equation \cref{eq: True dynamics integration} at the end of the segment are compared to the computed states obtained from the SQP solver, $x_{k+1}$. Equation \ref{eq: quadrature constraint} is referred to as quadrature constraints and forms part of the equality constraints in the OCP.
\begin{equation}
    {x}_{k+1} = x_\mathrm{T}(t_{k+1})      
    \label{eq: quadrature constraint}
\end{equation}
To ensure continuity between segments and minimize the defects, the states at the end of one segment, $ x_{k^-}$, must be equal to the states at the beginning of the consecutive segment $ x_{k^+}$. This results in the following constraint:
\begin{equation}
  x_{k^-} = x_{k^+}
\end{equation}

\subsubsection{Speed Reduction Criterion}\label{speed subsection}
As adopted in 1972, Rule 6 of COLREGs by \cite{international2003colreg} dictates that all marine vessels must operate at safe respective speeds, such that emergency maneuvers can be performed without endangering the ship, other ships, or structures. For many ports worldwide, the maximum limit for ships operating within the vicinity of the harbor area, in calm sea conditions is about $ 10 $ knots. Further, as noted from  \cite{rachman2023experimental}'s work, a controlled approach to berth enhances a smooth transition between the two stages of berthing as well as reduces excessive acceleration and deceleration within the last stage of the berthing process.

\begin{figure}[htbp]
    \centering
   \includegraphics[width=\columnwidth]{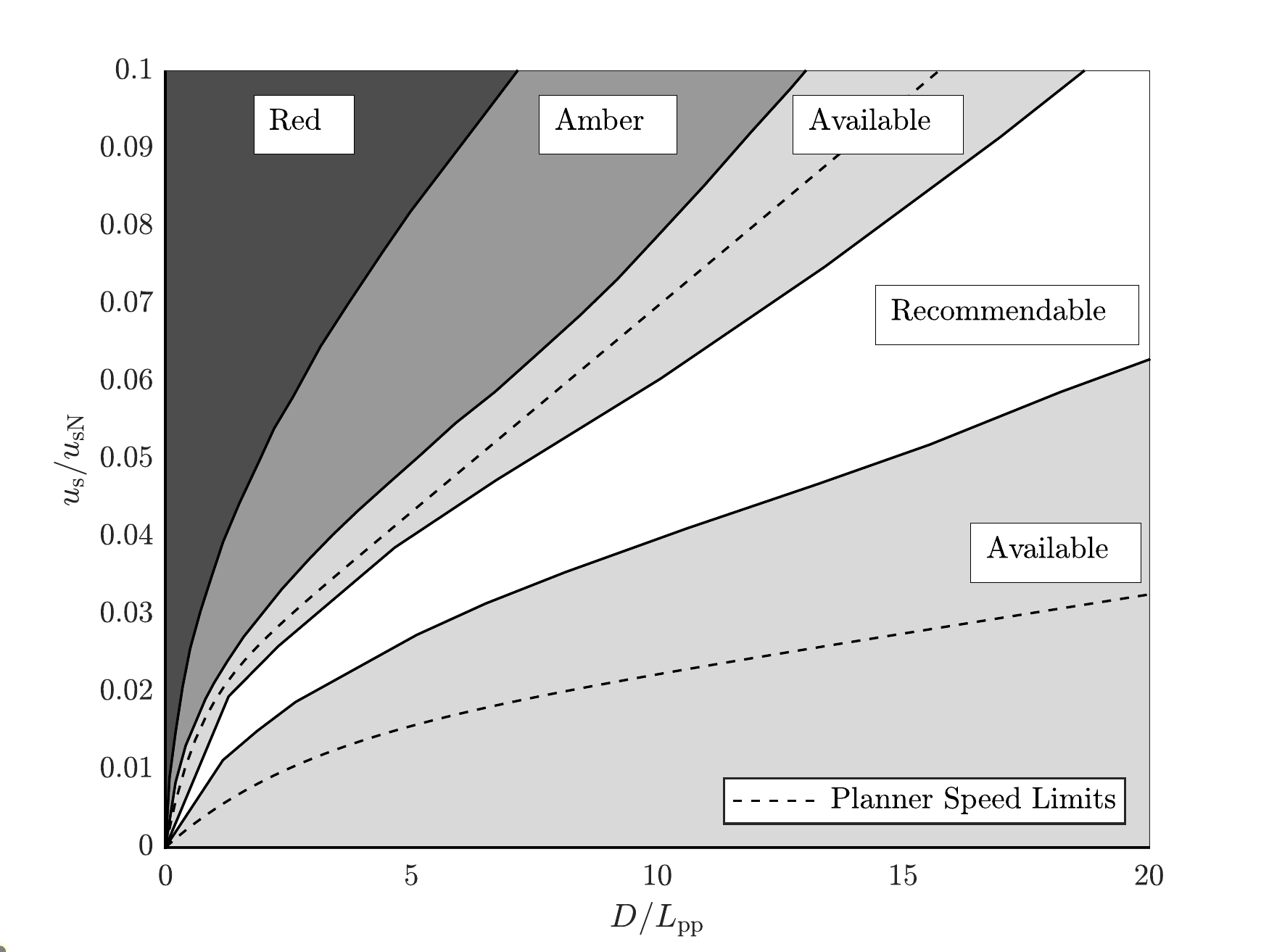}
    \caption{Speed Reduction criterion proposed by \cite{inouespeedreduction2002} }
    \label{fig: Inoue Speed criterion}
\end{figure}

\cite{KiyoshiHara1976} proposed a speed reduction model based on the operator's perception. From this model, it was clear that the speed reduction pattern is highly dependent on the individual operator and less dependent on ship size and port geometry. \cite{inoue1991assessment} proposed a safety margin-based model that relates the distance of the ship from the berth, ship speed, and the braking power. The safety margin is significantly low if the ship with low braking capacity approaches the speed at high speeds.\\
 \cite{inouespeedreduction2002} developed guidelines for ship speed reduction for ship berthing based on data obtained from ship captains' experiences obtained through a questionnaire. The guidelines developed consider both operational safety and captains' perceptions of safety, where each region represents a certain safety margin. As shown in \cref{fig: Inoue Speed criterion}, within the '\textbf{Red}' region, even if Full Astern braking force is used, the ship cannot achieve zero speed before reaching the target berthing point. While it is possible to stop the ship using either Full Astern, Astern, or Slow Astern braking force, operating within the '\textbf{Amber}' region involves a high risk of losing control of the ship. Within the '\textbf{Available}' regions, the captain can use Dead Slow Astern or Slow Astern braking force and can also easily change the ship's course without the risk of losing control. From the questionnaire data, most captains operate within the '\textbf{Recommendable}' region with Dead Slow Astern braking force as it offers the highest safety margin.
 
 \begin{figure}[htbp]
    \centering
    \includegraphics[width=\columnwidth]{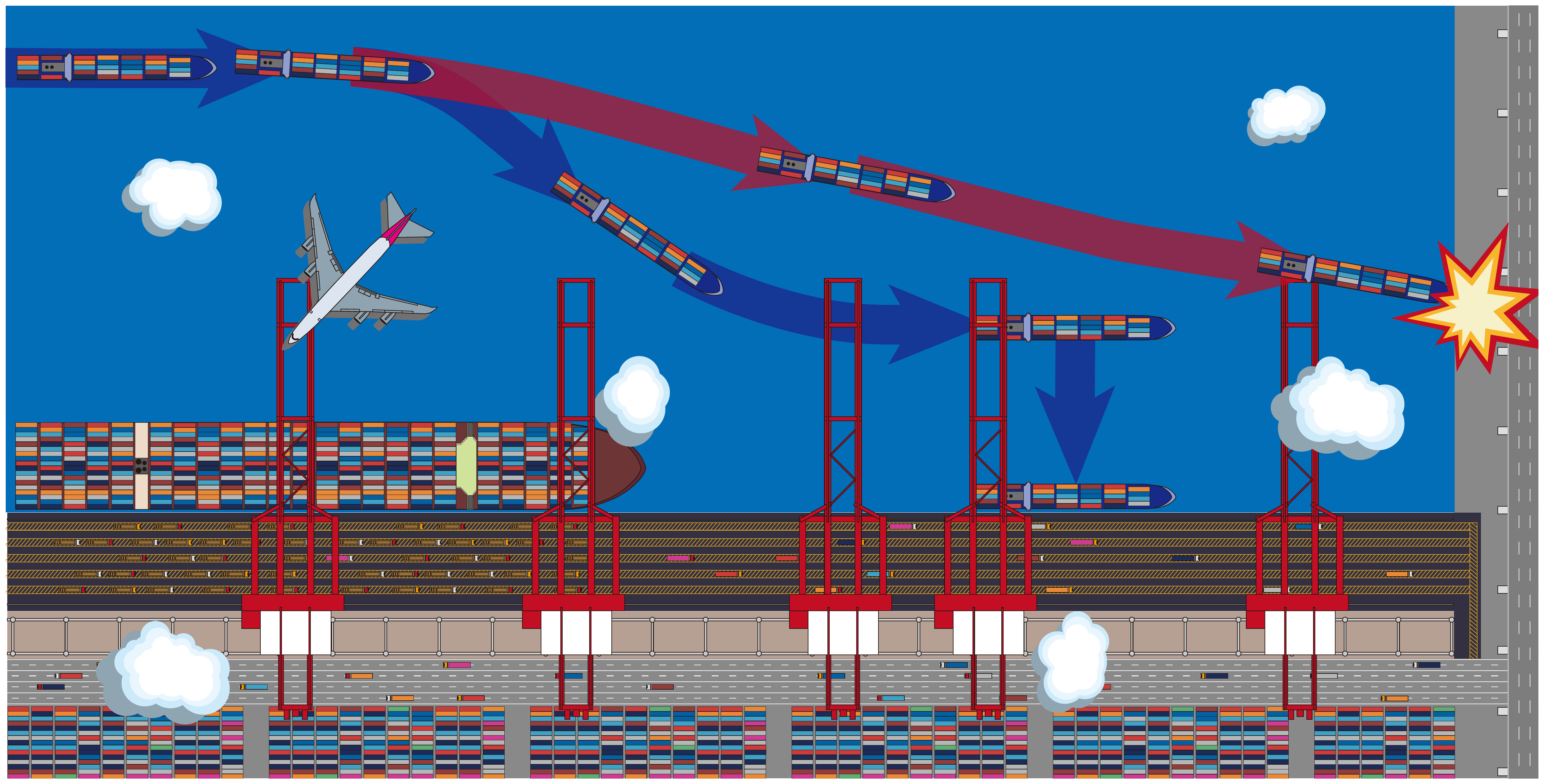}    
    \caption{This schematic image can be regarded as a summary of the speed reduction guidelines proposed by \cite{inouespeedreduction2002}. It depicts two berthing scenarios: (1) The ship approaches the berth with reduced speed (blue path) and reaches the berth safely. Along this path, the braking force may be as low as Dead Slow Astern or as high as Full Astern. (2) The ship approaches the berth with high speed (red path), and even with Full Astern as the braking force, the ship cannot stop at the desired berthing point and consequently collides with the pier, jetty, or wall.}    
    \label{fig: schematic speed reduction}
\end{figure}

 The proposed planner uses these guidelines to derive the speed reduction criterion, and setting the desired maximum speed limit is defined as shown in \cref{fig: Inoue Speed criterion}. $u_{\mathrm{s}}$, $u_{\mathrm{sN}}$, $D$ denote the ship's forward speed in $\mathrm{m/s}$, nominal ship speed in $\mathrm{m/s}$, and ship's distance from the berth in $\mathrm{meters}$, respectively.
 
Using non-linear regression, equation (\ref{eq:non-linear reg}) is formulated to determine the desired minimum and maximum speed limit with respect to the ship's distance from the berth.
\begin{equation}
    u_\mathrm{d(min, max)} = c_{1} d + c_{2}( 1 - e^{-c_{3} d} ),
    \label{eq:non-linear reg}
\end{equation}
where $u_\mathrm{d}$ and $d$ denote the non-dimesional ship's forward speed and ship's distance from the final berthing point respectively, that is, $u_\mathrm{d} = \uvelo/u_{\mathrm{sN}} $ and $ d = D/ L_{\mathrm{pp}} $. The values of the coefficients $c_{1}, c_{2}$ and $c_{3}$ are shown in \cref{tab: coefficients of speed reduction criterion}.

\begin{table}[htbp]
    \centering
    \caption{Coefficients of the Speed Limits Equation}
    \label{tab: coefficients of speed reduction criterion}
    \begin{tabular}{c|c|c}
    \hline
    Coefficient & Lower Speed Limit & Upper Speed Limit\\
    \hline \hline
       $c_{1}$ & $1.50 \times 10^{-3}$  & $5.06 \times 10^{-3}$ \\   
       $c_{2}$  & $1.70 \times 10^{-2}$ & $2.04 \times 10^{-2}$\\ $c_{3}$ & $3.78 \times 10^{-1}$ & $1.10$  \\
       \hline
    \end{tabular}    
\end{table}

Equation (\ref{eq:non-linear reg}) forms part of the inequality constraints in the final OCP, such that, for all knot points,
\begin{equation}
     u_\mathrm{d (min)}(t_k)\; \leq \; \uvelo (t_k) \;  \leq \; u_\mathrm{d (max)}(t_k)
\end{equation}

\subsubsection{Collision Avoidance Constraints} \label{sec: collision avoidance constraints}
The point-in-polygon (PIP) method is one of the simplest methods for determining the spatial characteristics of an object with respect to another object. Consider a 2D polygon with vertices $\mathbf{P}_0, \mathbf{P}_1, \mathbf{P}_2,\dots,\mathbf{P}_{N}$. The conditions $N \geq 3$ and $ \mathbf{P}_0 = \mathbf{P}_N$ must be satisfied for the polygon to be closed. There are two standard methods used to determine if a point lies inside a polygon, namely, the crossing number (odd-even rule) method and the winding number method.

The crossing number method dictates that a point $ \mathrm{Q} $ lies inside a polygon if a straight line drawn from the point to a point $\mathrm{S}$ outside the polygon crosses an odd number of edges, and $ \mathrm{Q} $ lies outside the polygon if the line crosses an even number of edges. On the other hand, the winding number method involves counting the number of times the polygon winds around the point. Moreover, as presented by \cite{hormann2001point}, the winding number method can be summarized by finding the summation of (signed) angles Q subtends with $\mathbf{P}_{i}\mathbf{P}_{i+1}$ edges for $i = 1, 2,\dots, N-1$, such that the winding number $\mathbf{wn}( \mathrm{Q},\mathbf{P}) = 2\pi$  when $\mathrm{Q}$ lies inside the polygon and $\mathbf{wn}( \mathrm{Q},\mathbf{P}) < 2\pi$  when $\mathrm{Q}$ lies outside the polygon. When dealing with simple convex polygons, any of the two methods is applicable.

In this study, the winding number method is used to define the collision avoidance constraints by considering the port layout as a closed 2D planar polygon and the vertices of the ship domain as the points whose inclusion in the polygon is to be determined. The ship domain is the effective 2D region surrounding a ship that should be kept free with respect to other ships or port structures. Generally, the shape and size of the ship domain depend on factors such as ship dimensions and maneuverability, environmental conditions, and ship speed, among others. In this study, an ellipse-shaped ship domain whose size is dependent on the resultant ship speed as presented by \cite{MIYAUCHI2022110390} is used.

Let $N$ and $\mathbf{P}_{i}$ denote the number of port boundary vertices and the $i_\mathrm{th}$ vertex of the port boundary, respectively, such that $i = 1, \dots, N$. $N_{\mathrm{sd}}$ denotes the number of ship-domain vertices, and $\mathbf{Q}_j $ is the $ j^ \mathrm{th} $  vertex of the ship domain such that $j = 1,\dots, N_{\mathrm{sd}}$.Let $\theta_{i}$ be the angle subtended from the $j_\mathrm{th}$ vertex of the ship domain by two consecutive vertices of the port boundary; $\mathbf{P}_{i}$ and $\mathbf{P}_{i+1}$.

\begin{figure}[htbp]
    \centering
    \includegraphics[width=\columnwidth]{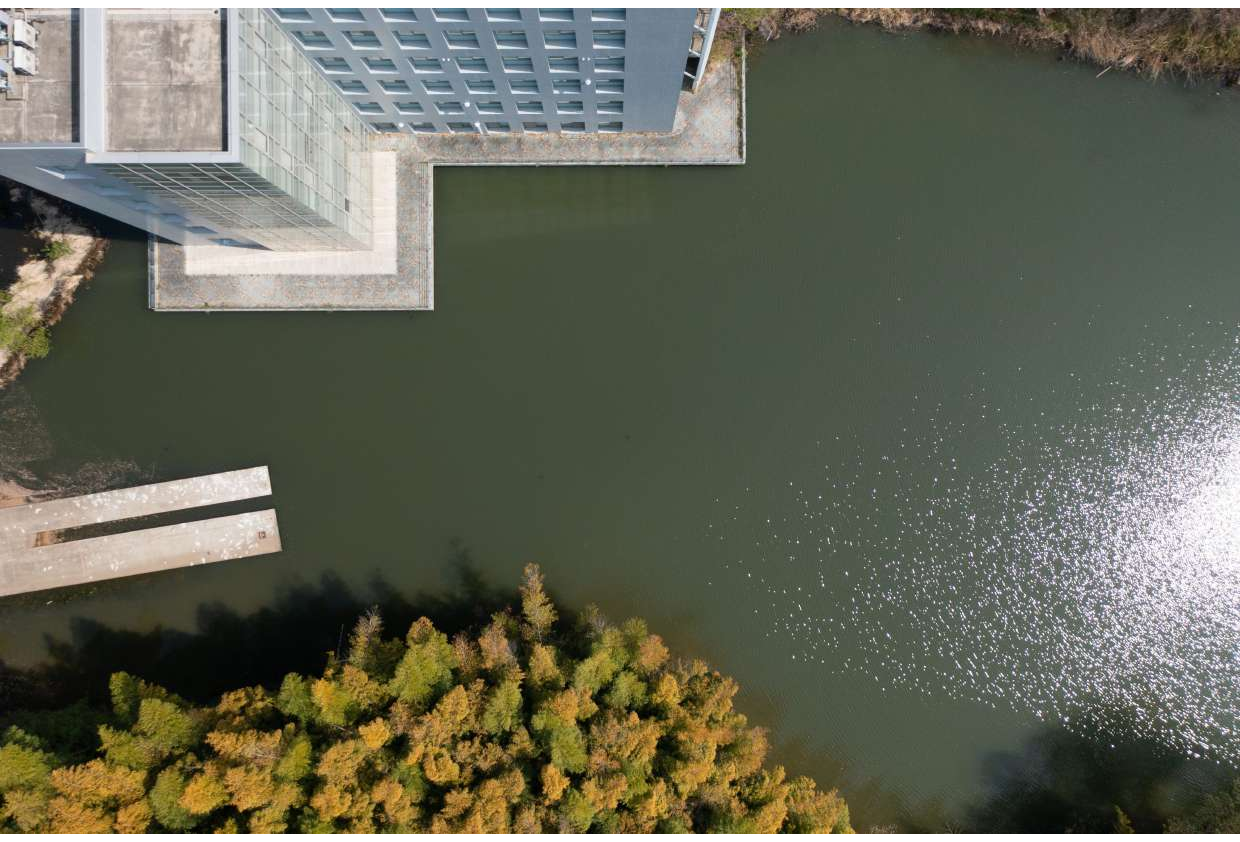}
    \caption{Inukai Pond, Osaka University}
    \label{ fig: Inukai Pond}
\end{figure}

\begin{figure}[htbp]
    \centering
    \includegraphics[width=\columnwidth]{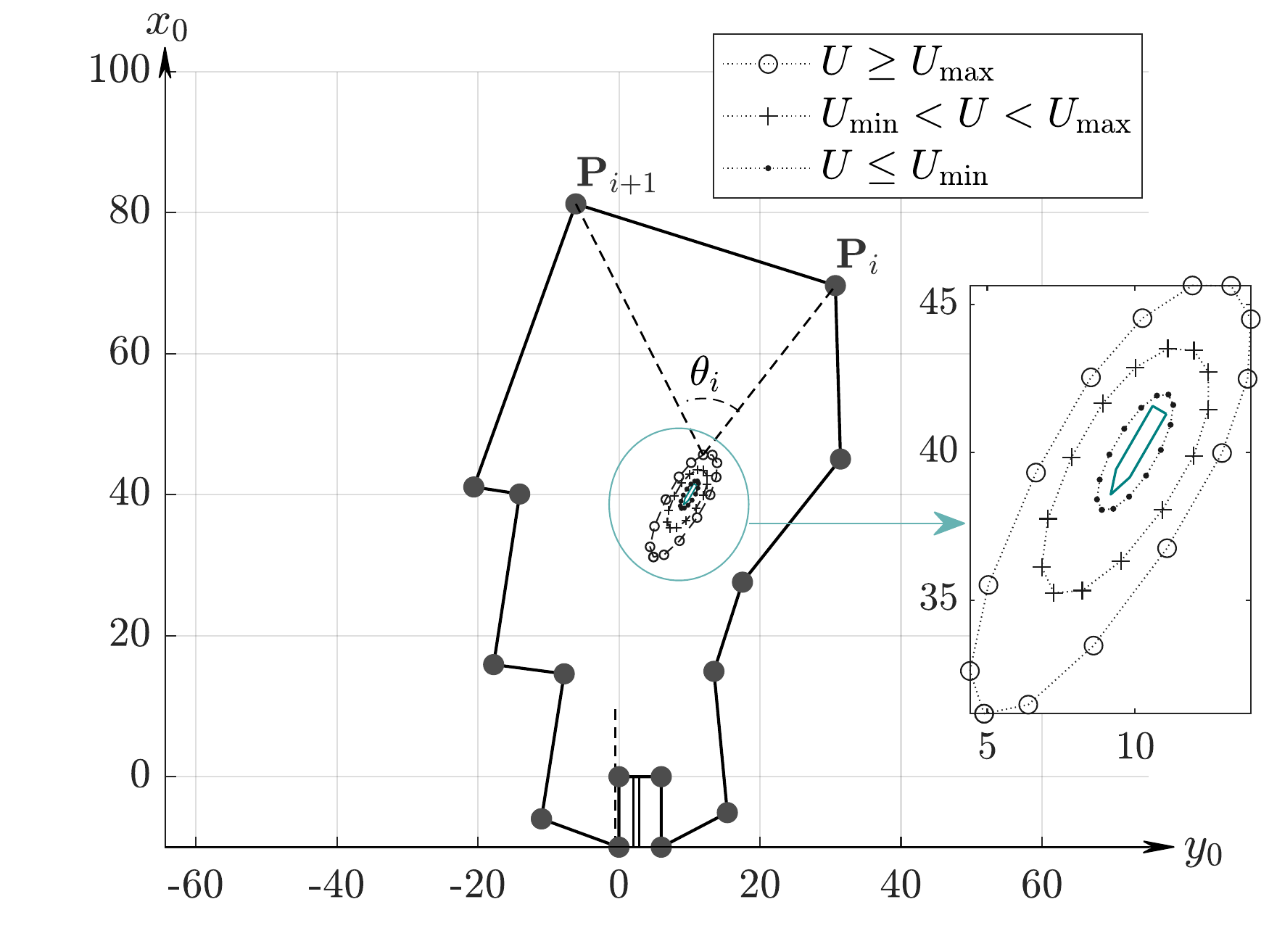}
    \caption{Port Boundary and Ship Domain Vertices}
    \label{ fig: collision avoidance constraints}
\end{figure}

The shape and location of port boundary vertices, as shown in \cref{ fig: collision avoidance constraints} are based on the topography of the Inukai Pond at Osaka University, as shown in \cref{ fig: Inukai Pond}. The position of these vertices is measured in the earth-fixed coordinate system, and the area inside the polygon is considered obstacle-free.

At every knot point, the sum of $\theta_{i}$ subtended from each ship domain vertex and all the port boundary vertices is computed as shown in \cref{eq: collision avoidance constraints}.
\begin{equation}\label{eq: collision avoidance constraints}
     \mathbf{wn}( \mathbf{Q}_j, \mathbf{P}) = \sum_{ i = 1}^{N - 1} \theta_{j,i}{\Big\vert}_{k} 
\end{equation}

Since the position of the ship domain vertices is dependent on the ship's yaw angle ($\psi$), velocities($\uvelo$ and $\vm$), and the midships position $(x_0, y_0)$, \cref{eq: collision avoidance constraints} forms part of the OCP constraints.

\subsubsection{OCP after Transcription}

Since the ship is operated with constant propeller revolutions, $n_{\text{p}}$ is excluded from the optimization variables. The vector of unknown variables, $\mathbf{X}$, to be optimized is expressed as:
\begin{equation}
    \mathbf{X}\; \equiv {(\tfpla, x_{1},\dots,x_{N_k}, u_{1},\dots,u_{N_k-1} )} ^ \intercal \in \mathbb{R}^{N_k + (N_k - 1) + 1}
\end{equation}   
The optimal control problem to be solved is finally defined as:\\
\begin{subequations}\label{eq:litdiff}
minimize:
    \begin{multline}        
        J = \sumstate  \norm{x_i \tf -x _{\mathrm{f}, i}}  ^ 2 \times  \int_{0}^{\tfpla} \sumstate \norm{x_i(t) - x_{\mathrm{f}, i}}  ^ 2 \dt,
    \end{multline}
subject to:
    \begin{eqnarray}
        \begin{cases}
           t_k \in [ 0, \tfpla]  \; \mathrm{ and } \; \tfpla \in 
           (0, \infty) \\[2pt]
           U_\mathrm{T}\tkpare = U_\mathrm{T}( \Bar{t}_{j}) \; \text{and} \; \gamma _\mathrm{T} \tkpare = \gamma_\mathrm{T}(\Bar{t}_j)\\[2pt]    
           x^*(t_0) = x_{0}\\[2pt]    
           x^*\tf = x_{\mathrm{f}}\\[2pt] 
           x_k^- = x_k^+ \; \text{for} \; k = 2,\dots,N_k \\ \notag       
           x_{k+1} = x_\mathrm{T}(t_{k+1}) \; \text{ for } \; k = 2,\dots,N_k \\[2pt]        
           u_\mathrm{d (min)}(t_k)\; \leq \; \uvelo (t_k) \;  \leq \; u_\mathrm{d (max)}(t_k) \; \mathrm{ for } \; k = 1,\dots,N_k\\[2pt]
           \sum\limits_{i=1}^{N-1} \theta_{j,i}{\Big\vert}_{k} = 2\pi \; \text{for} \; j = 1,\dots,N_\mathrm{sd} \; \mathrm{and} \;  k = 1,\dots,N_k\\[2pt]     
           u_{\mathrm{min}} \leq u(t_k) \leq u_{\mathrm{max}} \\[2pt]  
           n_{\mathrm{p}}(t_k) = 10  \text{rps} 
        \end{cases}
    \end{eqnarray}       
\end{subequations}
$ \tfpla, x_{0}, x_{\mathrm{f}}$ denote the final time, and desired initial and final states, respectively. $x^*(t_0), x^*\tf$ are the optimal initial and final states, respectively.
\subsection{Simulation Conditions}\label{sec : simulation }
\cref{tab: simulation conditions table} details some of the standard simulation conditions and computer specifications. 
\begin{table}[h!]
\caption{Simulation Conditions}
    \label{tab: simulation conditions table}
    \centering
    \setlength{\extrarowheight}{2pt}
    \begin{tabular}{l|p{40mm}}
    \hline \hline
       Distance from the berth &  $x_0(\text{f}) < D \leq 20\lpp$ \\
        Forward velocity & $\uvelo \leq 0.75 $ m/s. This corresponds to about 10 knots of full-scale ship.\\
        Computer used & 16GB RAM and Intel(R) Core(TM) i7-9700 CPU @ 3.00GHz Processor.\\
    \hline
    \end{tabular}   
\end{table}

\subsubsection{Harbor and Berth Approach Scenarios}
\begin{figure}[htbp]
    \centering
    \includegraphics[width=\columnwidth]{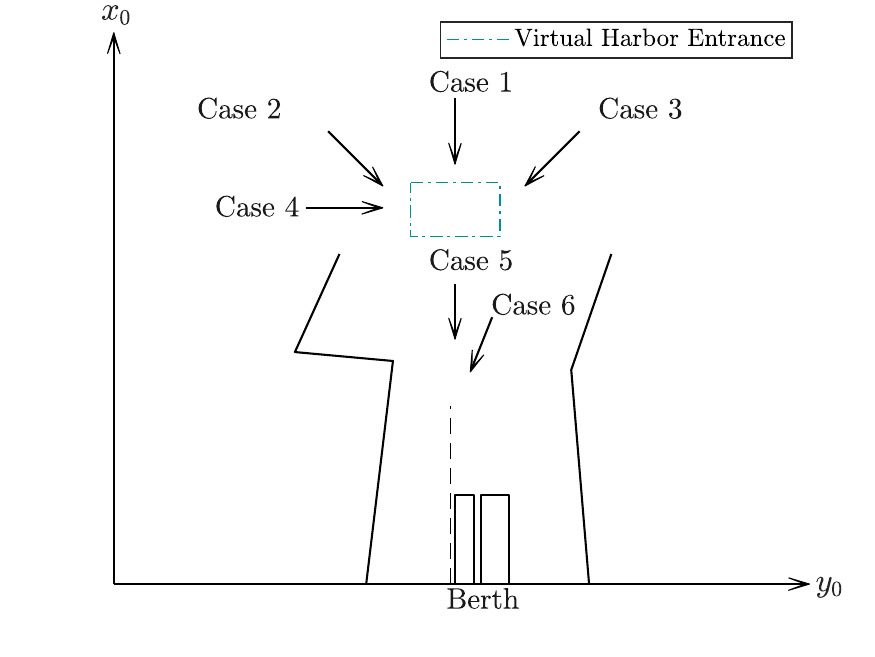}
    \caption{Simulation Cases}
    \label{fig: cases}
\end{figure}

By introducing a virtual harbor entrance, as illustrated in \cref{fig: cases}, practical harbor entry and berth approach scenarios were simulated without recomputation attempts as follows:
\begin{itemize}
    \item Case 1: Head-on approach to the entrance
    \item Case 2: Oblique approach to the entrance.
    \item Case 3: Similar to cases but on the opposite side of the berth. 
    \item Case 4: Parallel approach to the entrance, and perpendicular to the berth.
    \item Case 5: Past the harbor entrance, with an angular approach to the berth.
    \item Case 6: Past the harbor entrance, with a parallel approach to the berth.
\end{itemize}

\cref{tab: initial conditions} gives further details about Cases 1 - 6.

\begin{table}[h!]
\caption{Summary of Initial Conditions for the Simulation Cases}
    \centering
    \label{tab: initial conditions}
    \setlength{\extrarowheight}{2pt}
    
    \begin{tabular}{l|c c c c c c|c c}
    \hline 
    \multicolumn{1}{c|}{Case} & \multicolumn{6}{c|}{Initial States} & \multicolumn{2}{c}{Wind}\\
    \multicolumn{1}{c|}{} & \multicolumn{6}{c|}{ $ x(t_0)$} & \multicolumn{2}{c}{$(t = \Bar{t}_{j})$}\\
    \multicolumn{1}{c|}{} & $x_0$ & $\uvelo$ & $y_0$ & $v_m$ & $\psi$ & $r$ & $\gamma_\mathrm{T}$ & $U_\mathrm{T}$ \\
    \hline \hline
    1  & 60	&0.74   &0     &0	&3.14  &0  &0    &1.0  \\
    2  & 55.2 	&0.58	 &-6    &0	&2.36	&0  &45   &0.75  \\
    3  & 57.6	&0.58	 & 10    &0	&3.93	&0  &250  &0.5 \\
    4  & 52.8	&0.47	 &-10   &0	&1.57	&0  &45   &0.25 \\
    5  & 24.0   &0.34	 &6 	&0	&3.77	&0  &90   &0.5\\
    6  & 28.8   &0.29	 &0	    &0	&3.14	&0  &315  &0.75\\
    \hline
\end{tabular}
\end{table}
\subsubsection{Stochastic Generation of Initial Conditions and Positions}
For the feasibility study, trajectory planning is done for stochastically distributed initial positions and conditions with port boundaries. From one initial states vector, $ x_{\mathrm{i}}$, random initial states vectors, $x_{\mathrm{r}}$ are generated as follows:
\begin{equation}
 \begin{split}
    x &= {[\; 24, \; 0.29, \;-5, \;0, \;3.14, \;0\;]}^ \intercal\\    
    v_{\mathrm{r}} &= {[\; v_1, \;v_2, \;v_3, \;v_4, \;v_5, \;v_6 \;]}^ \intercal\\
    x_{\mathrm{r}} &= v_{\mathrm{r}} \odot x
 \end{split}
\end{equation}
$v_{\mathrm{r}} \in \mathbb{R} ^ 6 $ is a multiplier vector whose elements are stochastically generated as described in \cref{alg: random multiplier vector generator}.
\begin{algorithm}[ht!]
    \caption{Algorithm for generating $v_{\mathrm{r}} \in \mathbb{R} ^ 6$}\label{alg: random multiplier vector generator}
    \begin{algorithmic}[1]
        \State  $v_1 \in [\; 0.2 \; 3.0\; ]$ , $v_2\in [ \;0.2 \;2.54\;]$ , $v_3 \in [\;-6.0 \; 4.0\;]$, $v_4 \in [ \;0.1 \; 1.0\;]$ 
        
        \If{$v_1 \leq 1.0$ and $v_3 \geq 0.0$}
           \State $v_5 \in [ \;0.5 \; 0.85 \;]$ %\Comment{multiplier for $\psi$ for $x_0 \leq 24$ and $ y_0 \leq 0 $}
        \ElsIf{$ v_1 < 1.0$ and $v_3 < 0.0$}
           \State $v_5 \in [\;1.35 \; 1.50\;]$ %\Comment{\textit{multiplier for $\psi$ for $x_0 \leq 24$ and $ y_0 > 0 $}}
        \ElsIf{$v_1 > 1.0$ and $v_3 < 0.0$}
           \State $v_5 \in [\;1.0 \; 1.5\;]$ %\Comment{multiplier for $\psi$ for $x_0 > 24$ and $ y_0 > 0 $}
        \Else
           \State $v_5 \in [\;0.5 \; 1.0\;]$ %\Comment{multiplier for $\psi$ for $x_0 > 24$ and $ y_0 \leq 0 $}
        \EndIf
        
        \State $v_6 \in [\;0.1 \; 1.0\;]$ %\Comment{multiplier for $r$}
    \end{algorithmic}
\end{algorithm}
The yaw angle's multiplier element $ v_5 $ is conditionally defined to ensure a suitable heading from all positions. The initial positions and velocity are also checked to ensure that they lie within the port boundaries and speed limits, respectively as shown in \cref{alg: position and speed checking}.

\begin{algorithm}[ht!]
    \caption{Algorithm to ensure initial conditions are within the port boundaries and speed limits bounds.}
    \label{alg: position and speed checking}
    \begin{algorithmic}[1]
        \State $\text{Compute the position vertices of the ship domain}$
        \State $\text{Solve point in polygon problem}$
        \If { $ \sum\limits_{i=1}^{N-1} \theta_{i,j}(t_0) = 2\pi \; \text{and} \; u_\mathrm{d (min)} \leq \uvelo(t_0) \leq u_\mathrm{d (max)}$}
            \State $ x(t_0) = x_{\mathrm{r}}$
        \Else
            \State $\text{Generate new } x_{\mathrm{r}}$
        \EndIf    
    \end{algorithmic}
\end{algorithm}

Further, the corresponding initial wind speed and direction are defined as shown in \cref{alg: wind}.
\begin{algorithm}[ht!] 
    \caption{Initial Wind Direction and Speed}
    \label{alg: wind}
    \begin{algorithmic}[1]
        \State $\gamma_{\mathrm{T}} \in [ \;0^\circ\; 360^\circ\;]$  
        \State $ U_{\mathrm{T}} \in [ \;0.0\; 1.0\;]$ 
    \end{algorithmic}
\end{algorithm}

\section{Results and Discussion}
\subsection{Algorithm Validation}
\begin{figure*}[H]
    \begin{subfigure}[b]{1.0\textwidth}
    \centering
   \includegraphics[width=\textwidth]{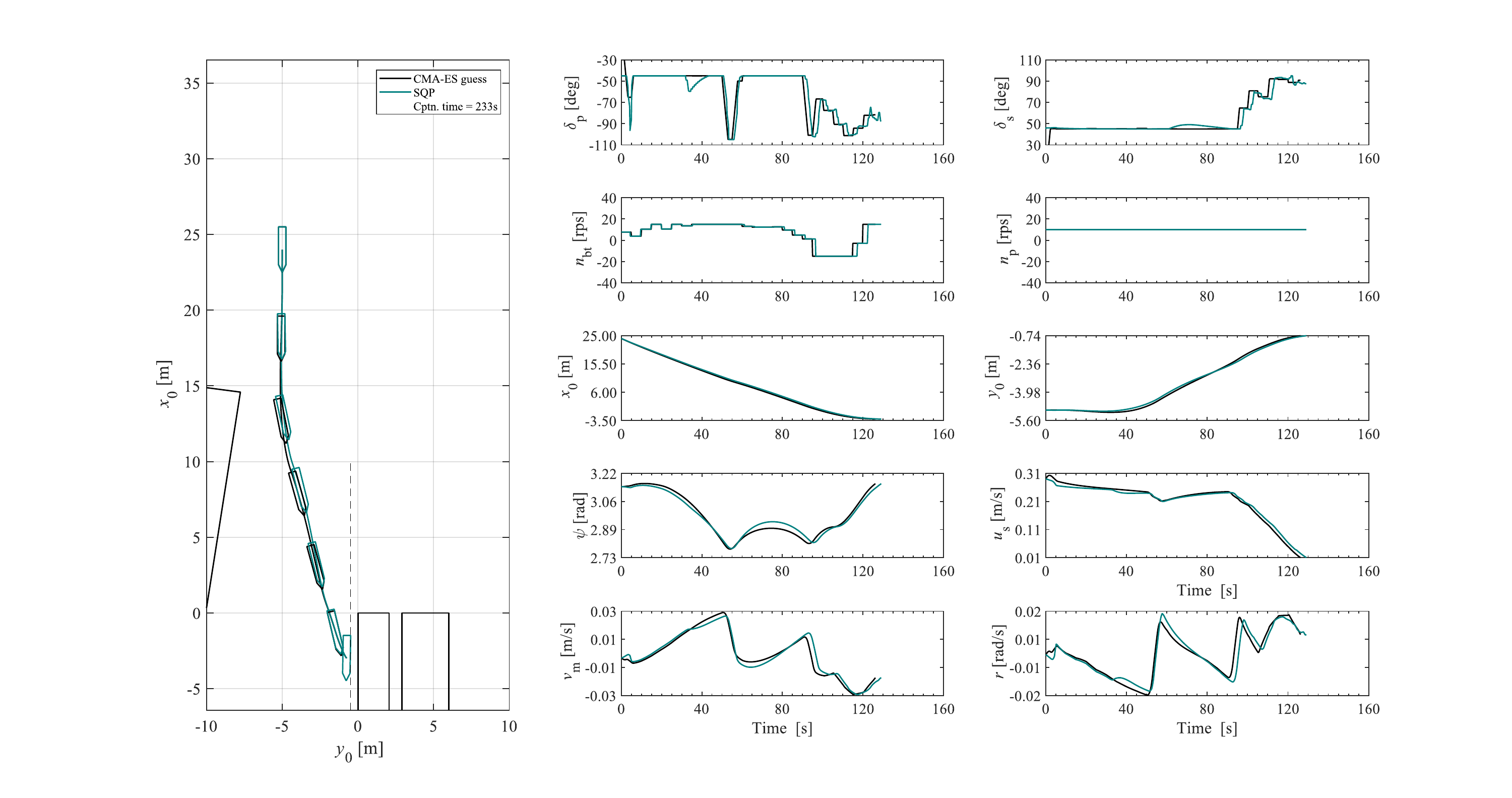}
    \caption{ }
    \label{fig: ship A}
\end{subfigure}
\begin{subfigure}[b]{1.0\textwidth}
    \centering
   \includegraphics[width=\textwidth]{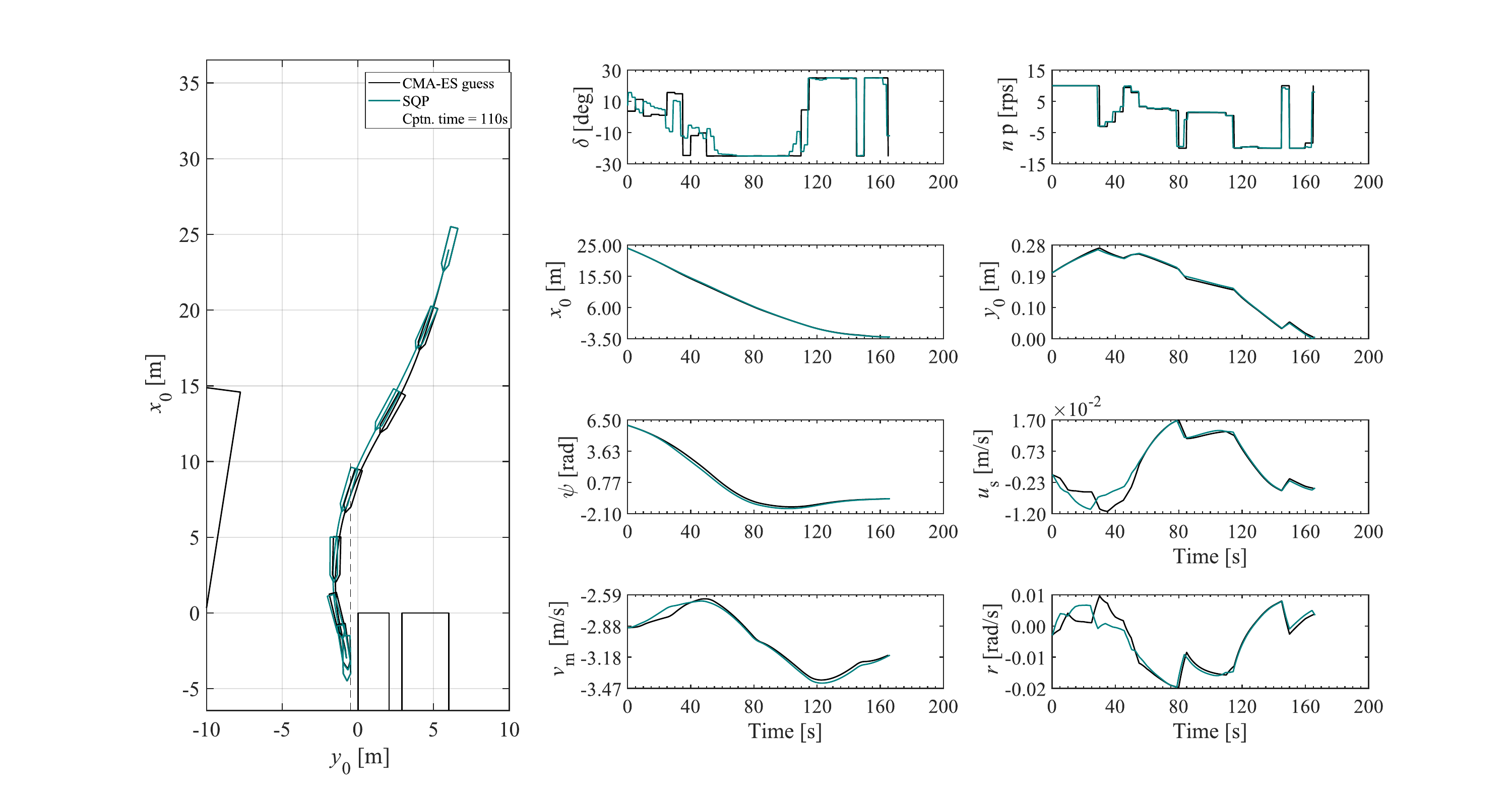}
\caption{ }
\label{fig: ship B}
\end{subfigure}
\caption{Optimal control inputs and states for \subref{fig: ship A} ship A  and \subref{fig: ship B} ship B when a solution from CMA-ES is used as the initial guess for the SQP computation. Wind disturbances and speed reduction are not considered.}
\end{figure*}

The accuracy and reliability of the proposed planner are demonstrated through comparison with optimal trajectories obtained using CMA-ES presented by \cite{maki2021application}, for two different model ships. Simulation is done without recomputation attempts and the speed reduction constraints are excluded from the simulation. Besides using the CMA-ES solution as the initial guess, the initial and final states $x(t_0)$ and $x(t_\text{f})$ as well as the wind conditions for the SQP are set similar to those of the CMA-ES.  Although their nature and approach differ, \cref{fig: ship A} and \cref{fig: ship B} demonstrate that optimal trajectories obtained using the proposed planner are almost identical to those obtained using CMA-ES, with the advantage of reduced computation cost. 

\subsection{Simulation Cases}
\subsubsection{Case 1: Head-on Approach to Harbor Entrance}
\begin{figure*}[htbp]
\centering
\includegraphics[width=\textwidth]{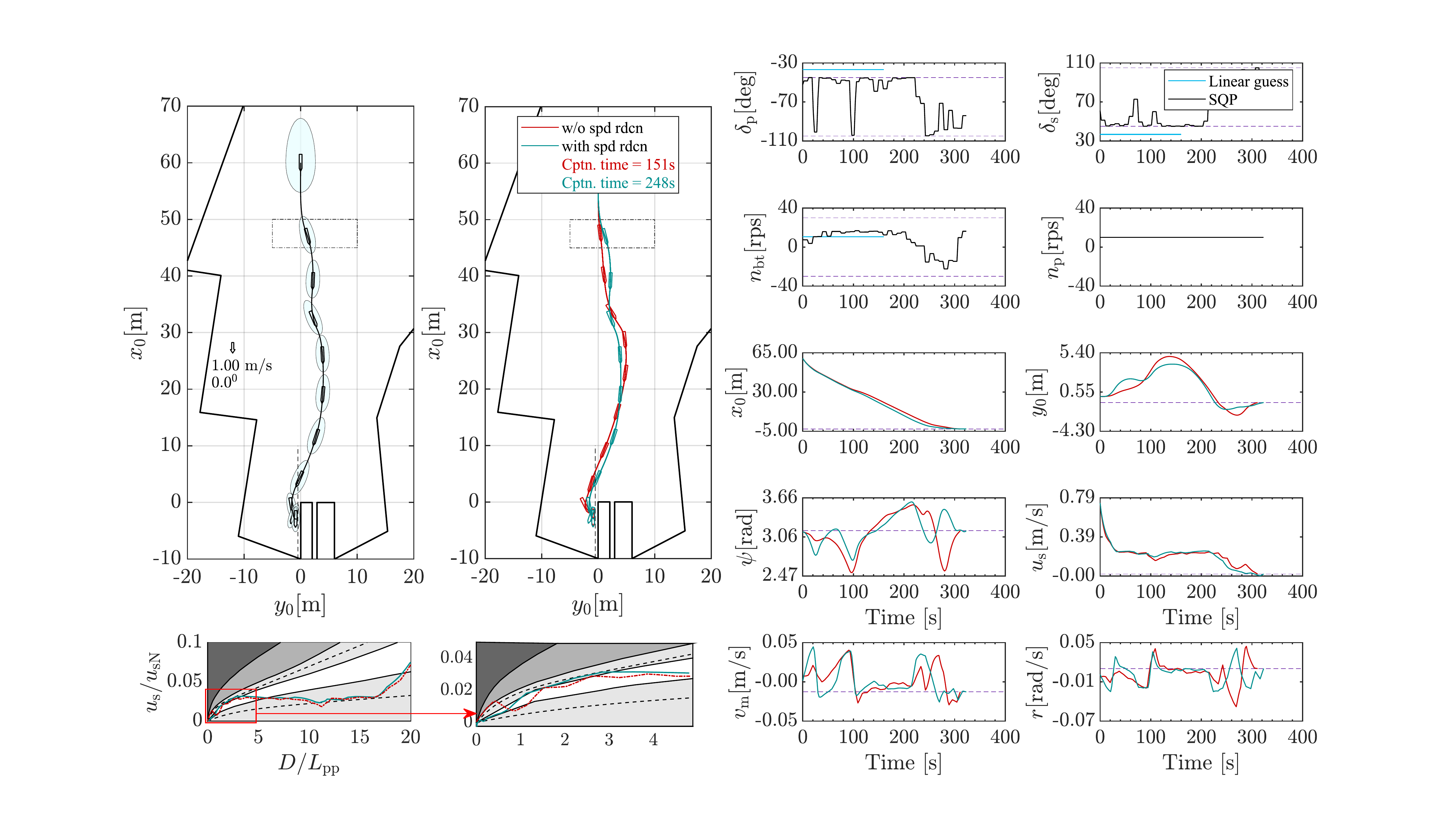}
\caption{ Case 1: Optimal trajectory, control, and states for a Head-On approach to the harbor entrance in the presence of wind disturbances, that is, $U_\mathrm{T} = 1.0 [\mathrm{ m/s}], \gamma_\mathrm{T} = 0.0^{\circ}$. The blue plots depict solutions incorporating speed reduction, whereas the red plots depict solutions without speed reduction.}
\label{fig: case1}
\end{figure*}

As shown in \cref{fig: case1}, the ship approaches a harbor or port directly from the open sea, moving in a direction that aligns its bow with the entrance of the harbor, such that the ship's longitudinal axis is perpendicular to the entrance. This approach is suitable during calm conditions for a ship with great course-keeping ability.

\subsubsection{Case 2: Oblique Approach to Harbor Entrance}
\begin{figure*}[H]
   \centering
  \includegraphics[width=\textwidth]{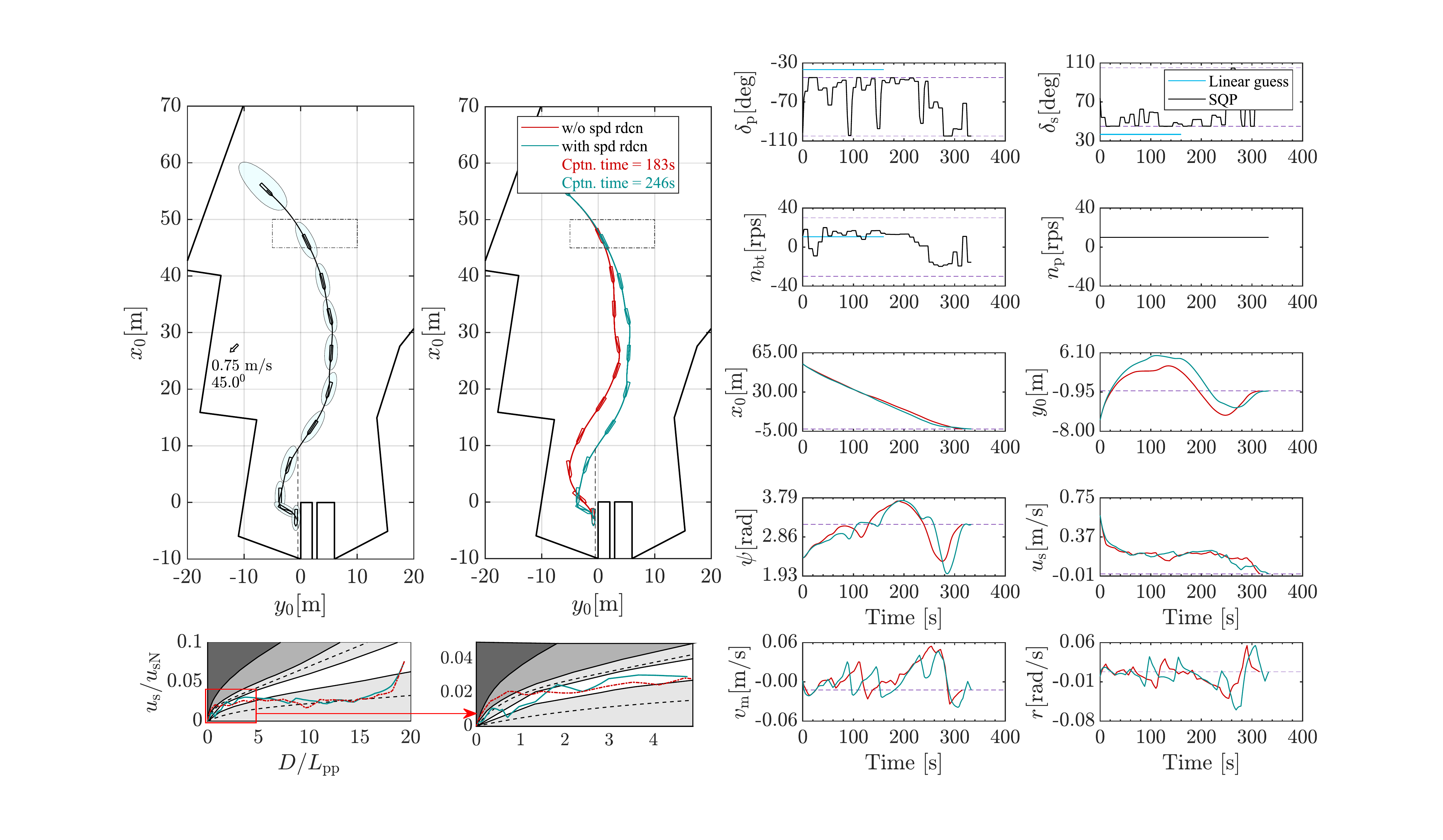}
\caption{Case 2: Optimal trajectory, control, and states for an Oblique approach to the harbor entrance in the presence of wind disturbances, that is, $U_\mathrm{T} = 0.5 [\mathrm{ m/s}], \gamma_\mathrm{T} = 45.0^{\circ}$. The blue plots depict solutions incorporating speed reduction, whereas the red plots depict solutions without speed reduction.}
\label{fig: case2}
\end{figure*}

In this scenario, as shown in \cref{fig: case2}, the ship is not directly aligned with the entrance but is approaching at an angle; that is, the ship's course forms an angle with the imaginary line that runs perpendicular to the harbor entrance.
As explained by \cite{national1981problems}, this approach helps counteract the impact of wind and strong currents and better manage the ship's stability in challenging weather conditions.

\subsubsection{Case 3: Oblique Approach to Harbor Entrance} 
\begin{figure*}[H]
  \centering
    \includegraphics[width=\textwidth]{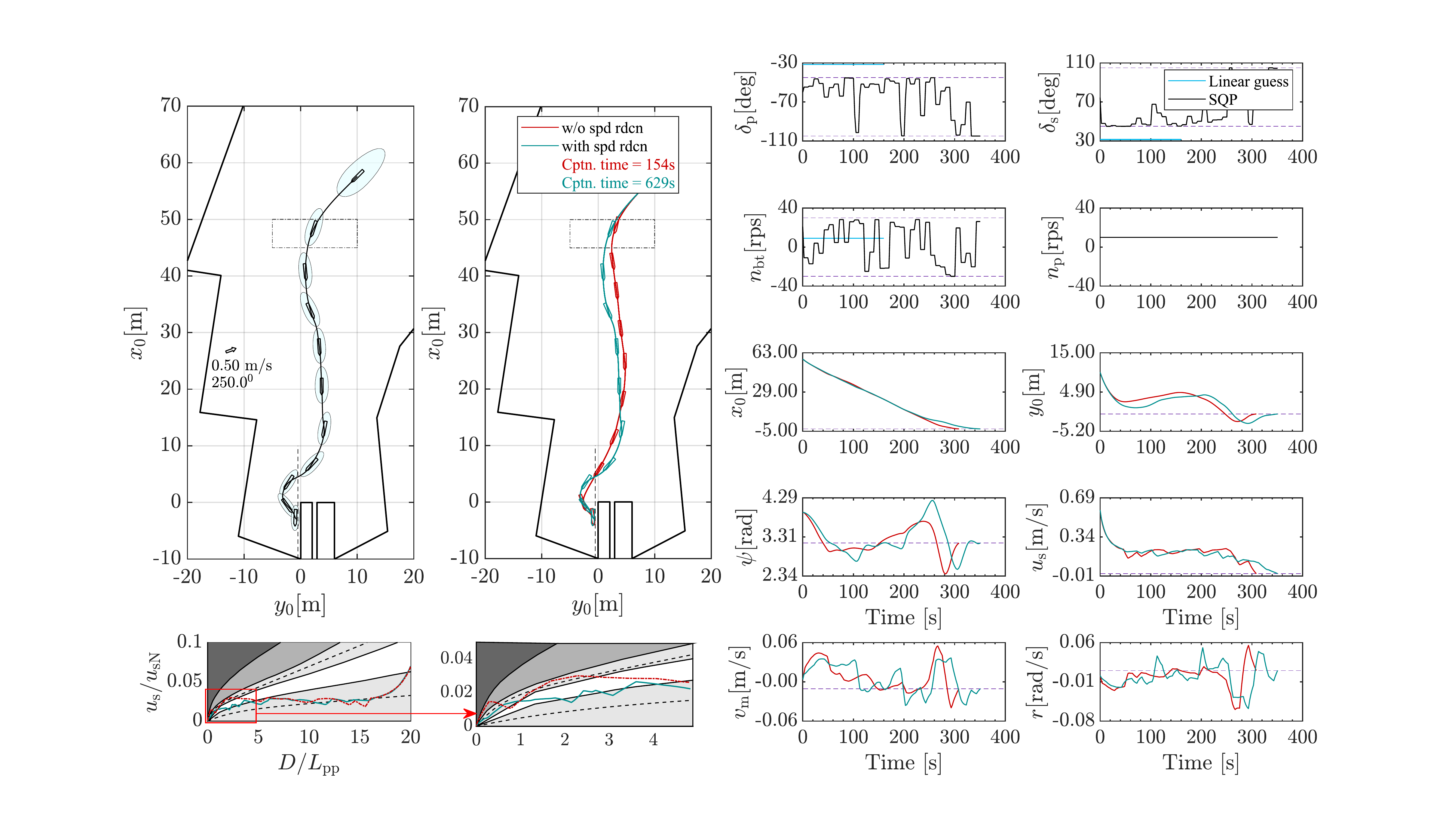}  
\caption{ Case 3: Optimal trajectory, control, and states for an Oblique approach to the harbor entrance in the presence of wind disturbances, that is, $U_\mathrm{T} = 0.5 [\mathrm{ m/s}], \gamma_\mathrm{T} = 250^{\circ}$. The blue plots depict solutions incorporating speed reduction, whereas the red plots depict solutions without speed reduction.}
\label{fig: case3}
\end{figure*}

This scenario is similar to case 2, but from the opposite side of the harbor entrance, as shown in \cref{fig: case3}. Further, as explained by \cite{national1981problems} this approach helps reduce the risk of accidents or difficulties when entering the harbor.

\subsubsection{Case 4: Parallel Approach to Harbor Entrance} 
\begin{figure*}[H]
  \centering
    \includegraphics[width=\textwidth]{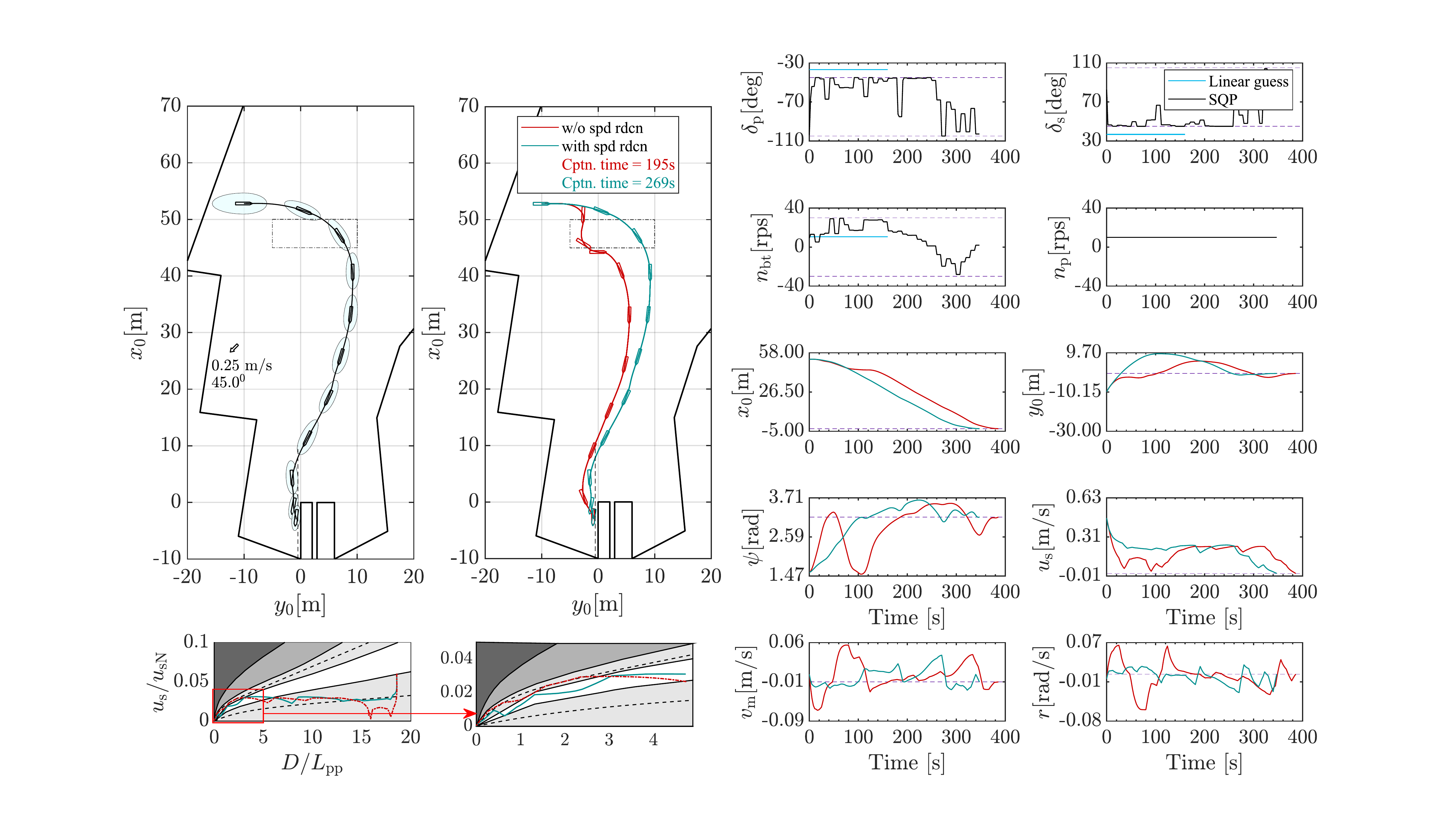}
\caption{ Case 4: Optimal trajectory, control, and states for a Parallel Approach to the harbor entrance in the presence of wind disturbances, that is, $U_\mathrm{T} = 0.25 [\mathrm{ m/s}], \gamma_\mathrm{T} = 45^{\circ}$. The blue plots depict solutions incorporating speed reduction, whereas the red plots depict solutions without speed reduction.}
\label{fig: case4}
\end{figure*}

As shown in \cref{fig: case4}, the ship's orientation with respect to the harbor entrance is such that its longitudinal axis is roughly parallel to the direction of the harbor entrance. This type of approach is common with large ships entering a narrow or constrained harbor.

\subsubsection{Case 5: Parallel Approach to the Berth}
\begin{figure*}[H]
    \centering
    \includegraphics[width=\textwidth]{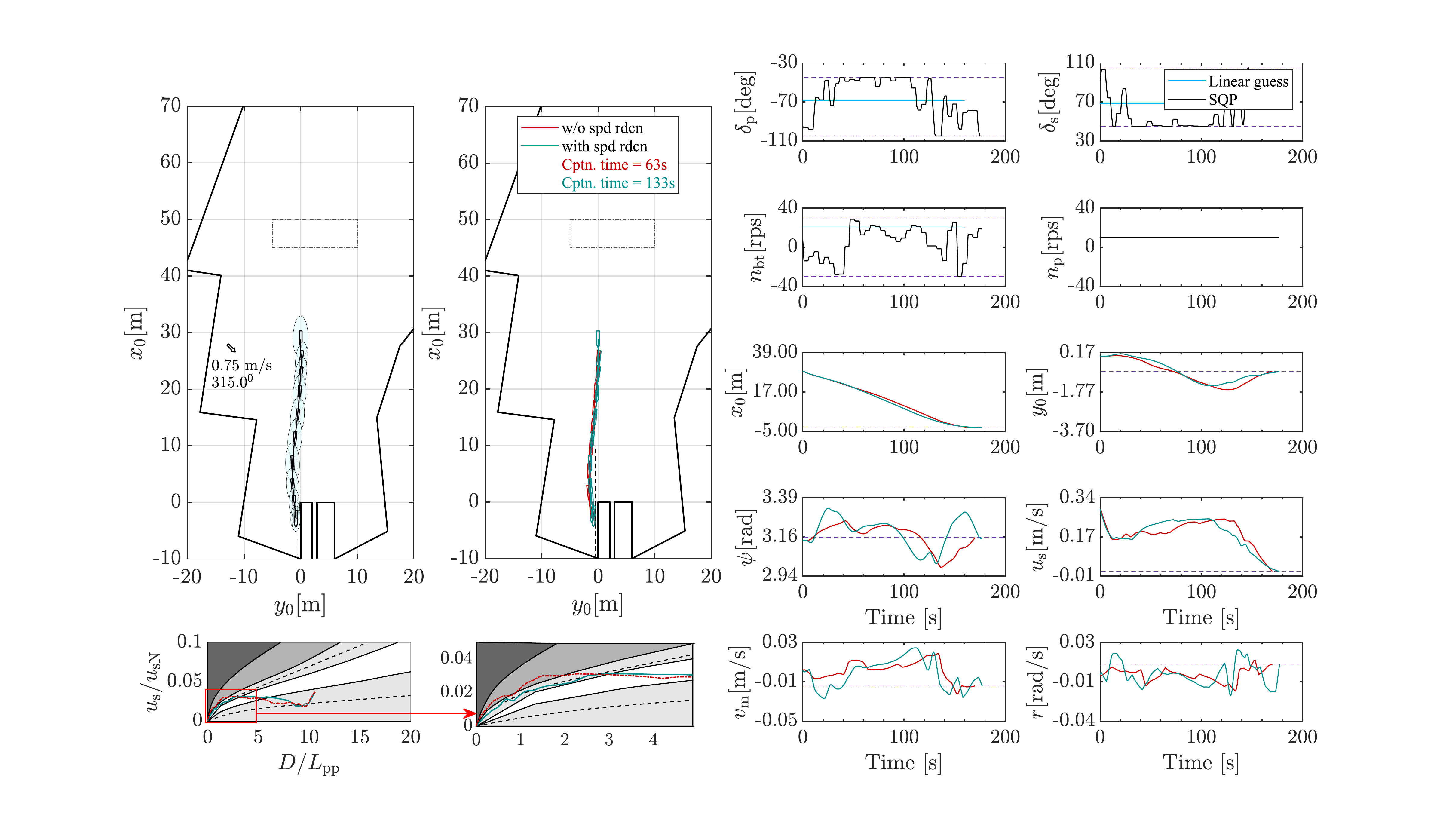}
\caption{ Case 5: Optimal trajectory, control, and states for a Parallel Approach to the berth in the presence of wind disturbances, that is, $U_\mathrm{T} = 0.75 [\mathrm{ m/s}], \gamma_\mathrm{T} = 315^{\circ}$. The blue plots depict solutions incorporating speed reduction, whereas the red plots depict solutions without speed reduction.}
\label{fig: case5}
\end{figure*}

Parallel approach to the berth as shown in \cref{fig: case5}, refers to the scenario where the ship approaches the berth in a way that its longitudinal axis is aligned with the berth's axis. This approach is suitable for a ship with great course-keeping ability and less crowded harbor areas.

\subsubsection{Case 6: Angular Approach to the Berth} 
\begin{figure*}[H]
    \centering
   \includegraphics[width=\textwidth]{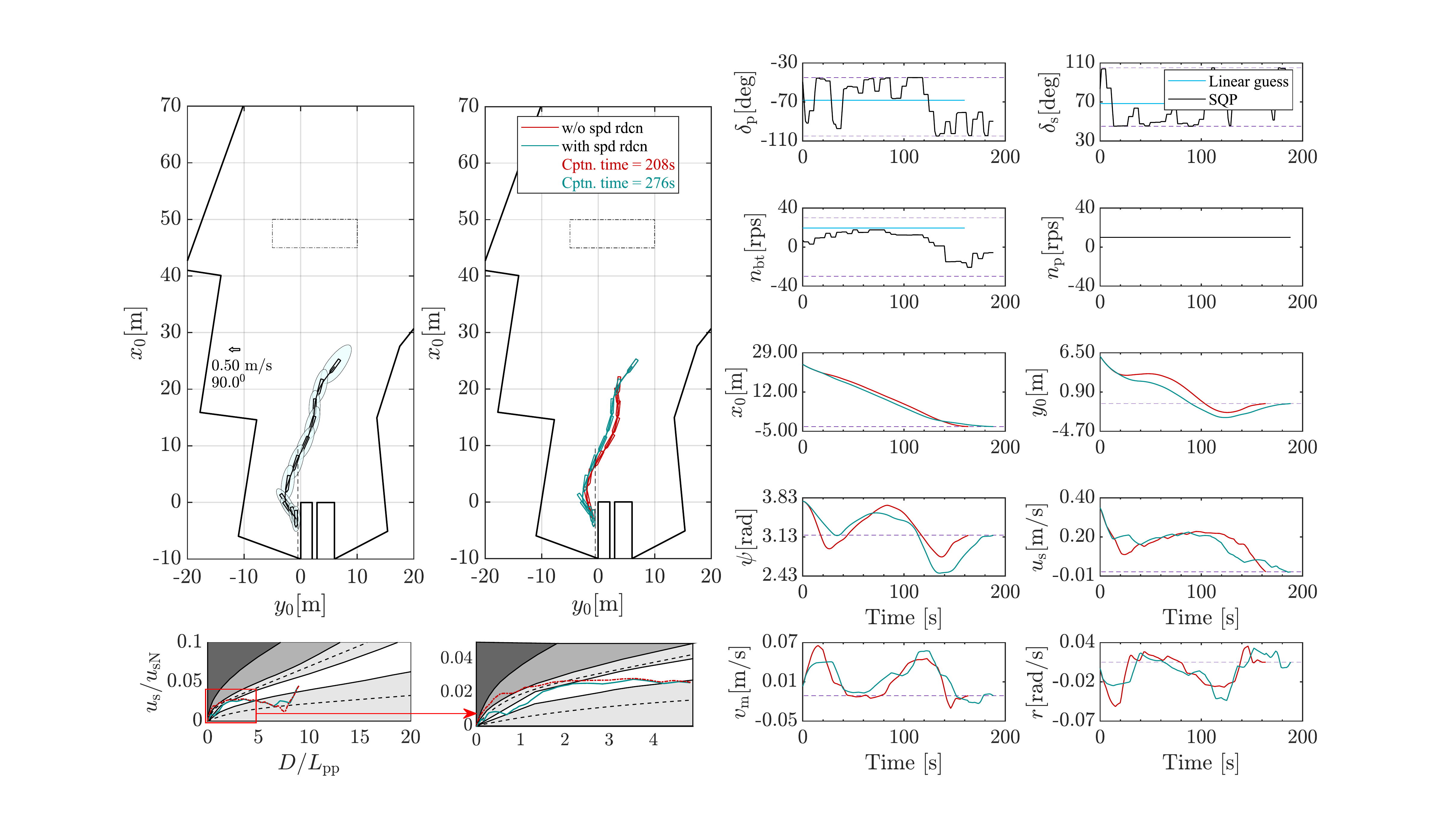}
\caption{ Case 6: Optimal trajectory, control, and states for an Angular Approach to the berth in the presence of wind disturbances, that is, $U_\mathrm{T} = 0.5 [\mathrm{ m/s}], \gamma_\mathrm{T} = 90^{\circ}$. The blue plots depict solutions incorporating speed reduction, whereas the red plots depict solutions without speed reduction.}
\label{fig: case6}
\end{figure*}

In this scenario, as shown in \cref{fig: case6}, the ship's orientation with respect to the berth is such that its longitudinal axis is at an angle with the berth. This approach is common in the presence of currents and winds or with large ships operating in confined harbor areas.

\subsection{Feasibility Study}
In the SQP context, "feasible" refers to a solution that satisfies all constraints of the optimization problem. On the other hand, "infeasible" describes the situation where the SQP algorithm fails to find a solution that satisfies all constraints. Sometimes, with a predefined number of maximum iterations, the solver may stop prematurely. In this study, cases where the solver stopped prematurely were categorized as infeasible. The feasibility study involved an assessment of 200 cases with stochastically generated initial conditions. Further, for the cases initially found infeasible, a maximum of three recomputation attempts were made with different control input initialization.

\begin{figure}[h!]
         \centering
         \includegraphics[width=0.9\columnwidth]{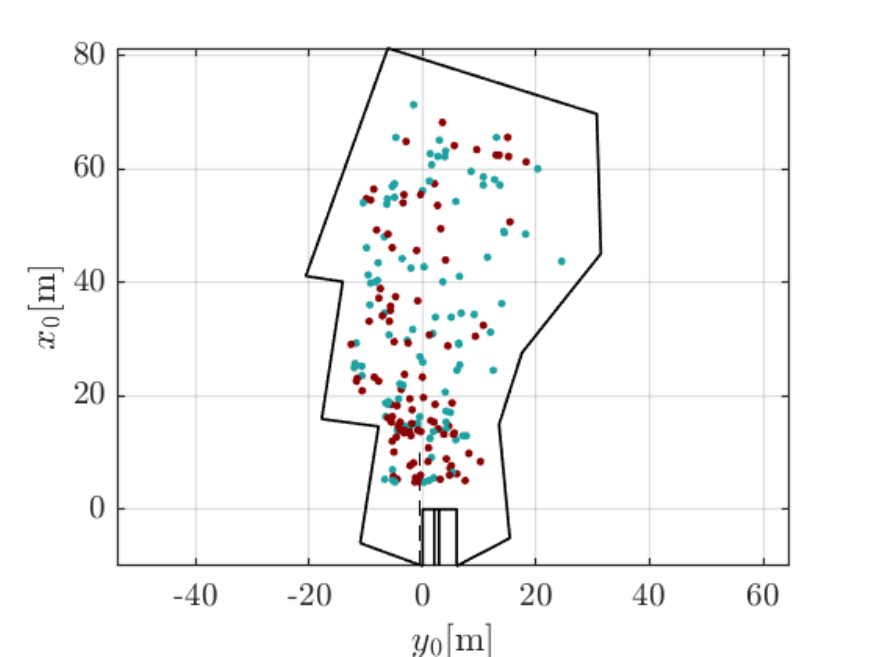}
         \caption{ Feasibility rate of the initial computation of which $ 50\%$ of the cases were feasible. Blue depicts feasible while red denotes infeasible.}
         \label{fig: feasibility1}
\end{figure}

\begin{figure}[h!]
         \centering
         \includegraphics[width=0.9\columnwidth]{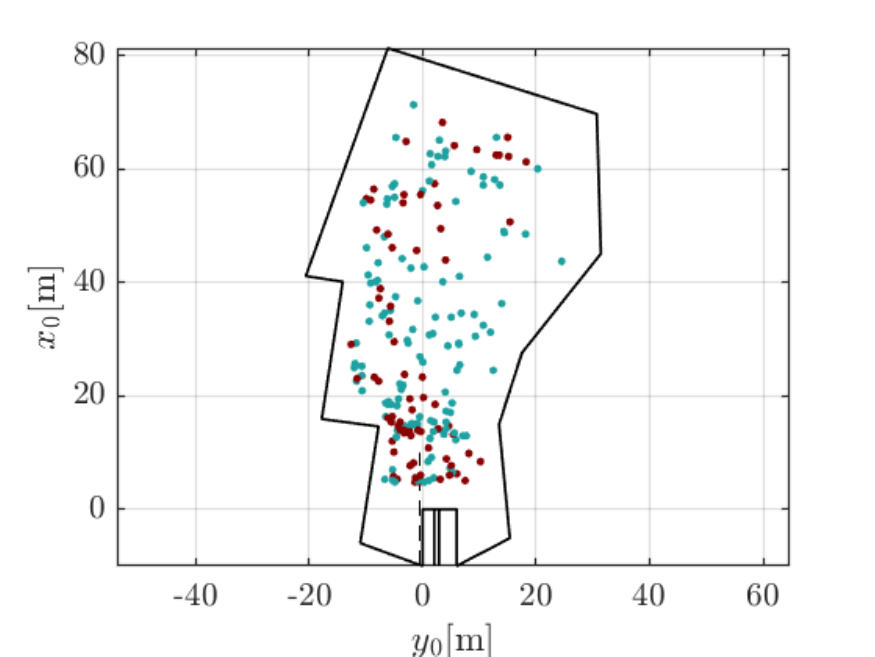}
         \caption{Feasibility rate of the first recomputation attempt. Blue depicts feasible while red denotes infeasible. The feasibility rate increased to $ 62.5\%$ }
         \label{fig: feasibility2}
\end{figure}

\begin{figure}[h!]
         \centering
         \includegraphics[width=0.9\columnwidth]{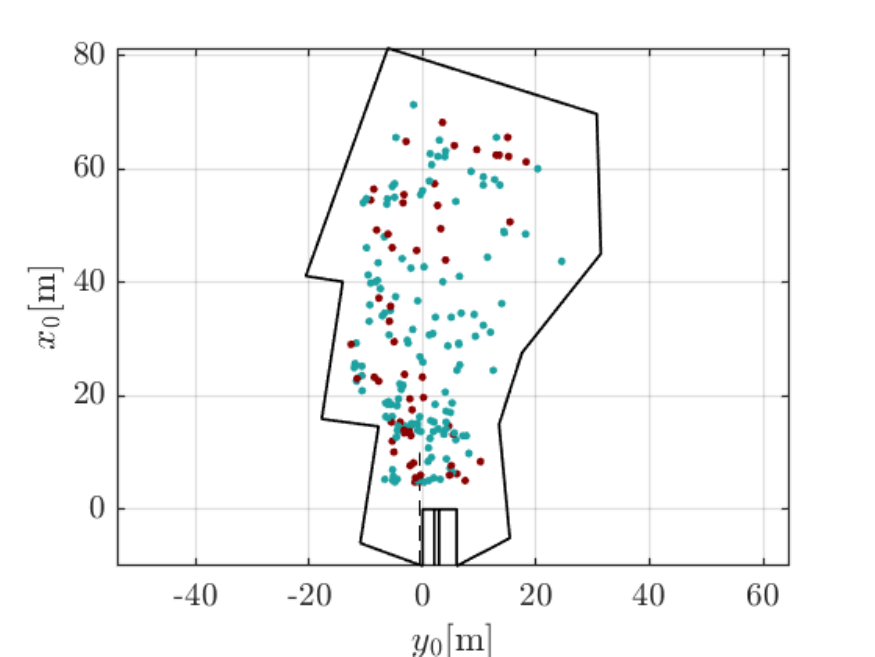}
          \caption{Feasibility rate of the second recomputation attempt. Blue depicts feasible while red denotes infeasible. The feasibility rate increased to $ 71\%$ }
         \label{fig: feasibility3}
\end{figure}

\begin{figure} [h!]
         \centering
         \includegraphics[width=0.9\columnwidth]{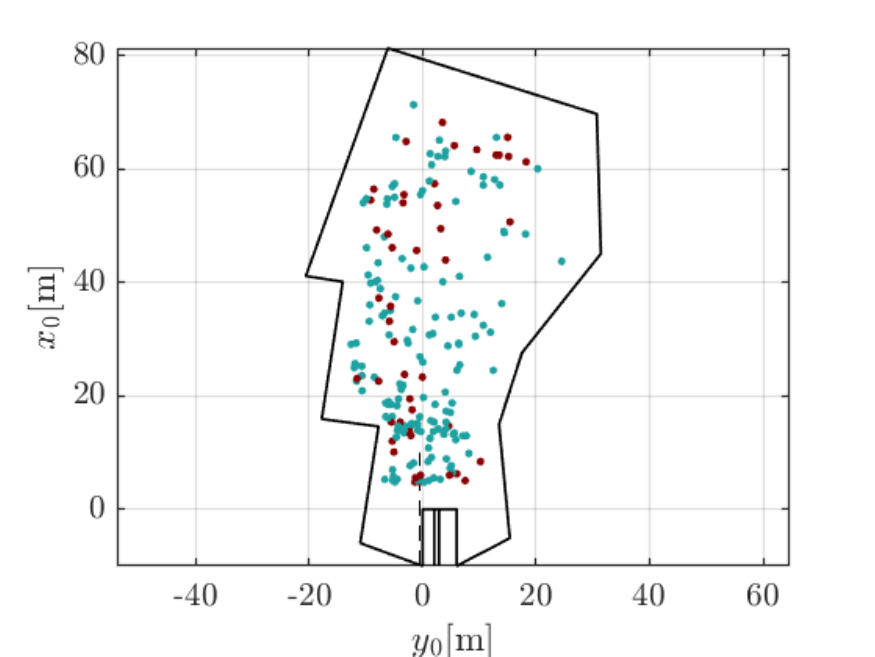}
          \caption{Feasibility rate of the third recomputation attempt. Blue depicts feasible while red denotes infeasible. The feasibility rate increased to $ 75\%$ }
         \label{fig: feasibility4}
\end{figure}

Initially, 100 cases ($50\%$) of the 200 cases under consideration were deemed feasible, as shown in \cref{fig: feasibility1}. \cref{fig: feasibility2} shows the results of the first recomputation attempt, where a total of 125 cases ($62.5\%$) were deemed feasible. \cref{fig: feasibility3}  and \cref{fig: feasibility4} show the results of the second and third recomputation attempts, where 142 cases ($71\%$) and 151 ($75\%$) cases were found feasible, respectively. 


\subsection{Computation Speed Analysis}
\begin{figure}[h!]
    \centering
    \includegraphics[width = 0.9\columnwidth]{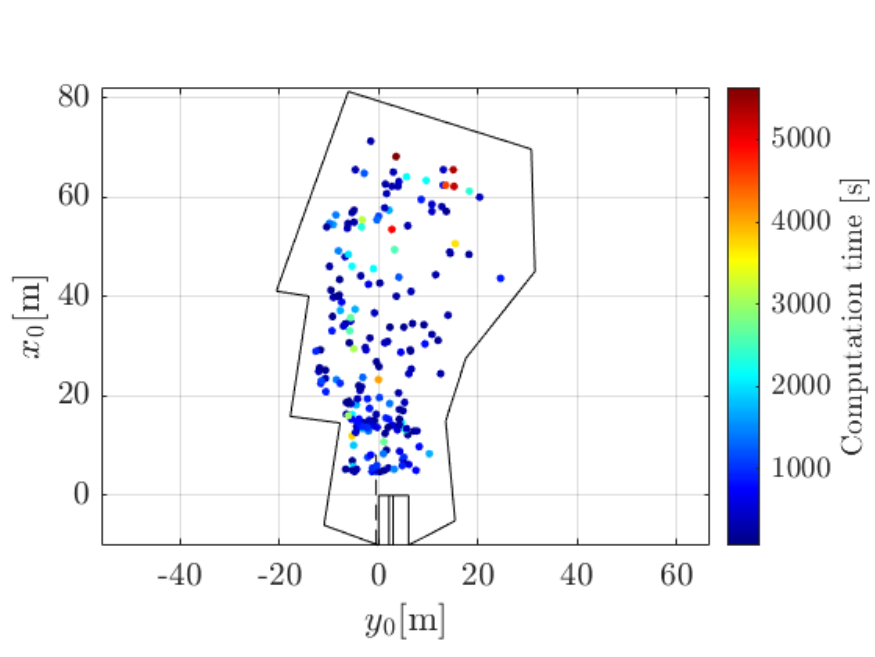}
    \caption{Summary of computation time for the final computation attempt for all 200 cases}
    \label{fig: computation time}
\end{figure}

\begin{figure}[h!]
    \centering
    \includegraphics[width = 0.9\columnwidth]{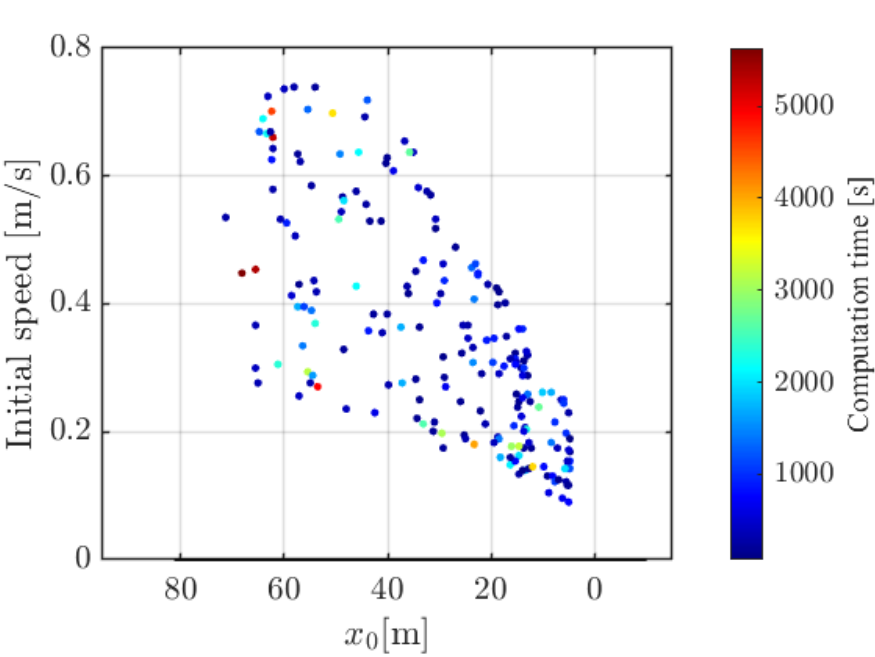}
    \caption{Summary of how  computation time varies with initial speed and distance from the berth}
    \label{fig: computation time2}
\end{figure}

\cref{fig: computation time} shows how the computation speed varies for the stochastically generated cases regardless of their feasibility status. It is noteworthy that the initial computation time was less than 300 seconds in 71 cases. However, after subsequent recomputation attempts, the total computation time per case increased, with the highest computation time being 5632 seconds. Moreover, \cref{fig: computation time2} shows that the proposed planner can generate physically and dynamically feasible solutions in as little as 69 seconds for a case closest to the berth and 274 seconds for a case furthest from the berth.

\section{Discussion and Limitations}
The proposed planner demonstrates a noteworthy departure from previous studies in the utilization of a simple linear guess to initialize the SQP algorithm. The simulation results in \cref{fig: case1} to \cref{fig: case6}, demonstrate the potential of using a less sophisticated initial guess for satisfactory performance of the trajectory planner. However, this has not been tested with a complex harbor environment, and/or in the presence of dynamic obstacles.

The inclusion of speed reduction and actuator limitations in NLP enhances the realism of the trajectory planner. From \cref{fig: case1} to \cref{fig: case6}, a comparison between solutions obtained without speed reduction and those without speed reduction indicates that in the vicinity of the berth, without speed reduction the ship approaches the berth within the $\textbf{'Amber'}$ region which poses a significant risk to ship handling and therefore its safety. Therefore, incorporating speed reduction constraints, which were not the focus of earlier research efforts is guaranteed to enhance berthing safety. Further, incorporating the actuator limitations creates a buffer margin that allows the system to counter unknown disturbances thereby enhancing traceability and operational safety.

From the feasibility study, initially, only $50\%$ of the cases were feasible. A notable improvement was achieved through three recomputation attempts for the initially infeasible cases, resulting in a $75\% $ feasibility rate. However, it was observed that changing only the control input guess during recomputation may not be sufficient to enhance overall feasibility. The infeasibility rate may be attributed to practically unrealistic initial positions and headings. Moreover, the prevailing wind speed and direction significantly impact the ship's heading, as noted by \cite{Wu2023}, and the ship's speed, as stipulated in \cite{guard2010new} guidelines. Subsequently, the wind speed and direction must be considered when defining the initial speed and heading.  Nonetheless, the results of the feasibility study highlight the robustness and effectiveness of the proposed planner in berthing trajectory optimization. It is noteworthy that although the computation time required for most cases was less than 300 seconds, the introduction of recomputation significantly escalated the computation time, reaching as high as 5000 seconds. This increase in the computational burden, while addressing infeasibility, may pose challenges in real-time applications and demand further investigation of the feasibility study.

%\clearpage
\section{Conclusion}
In summary, this study proposed a robust trajectory planner tailored for the autonomous berthing process, with a primary focus on enhancing safety. Besides meeting the regulatory requirements, the speed reduction criterion is guaranteed to enhance berthing safety and promote a controlled approach to the berth. Moreover, incorporating a rate of change and artificially limiting the actuators accounts for the physical constraints of the actuators, which reinforces the planner's viability in practical applications. Furthermore, the utilization of a ship domain as proposed by \cite{MIYAUCHI2022110390} in defining collision avoidance constraints increases the safety clearance margin, thereby significantly reducing the likelihood of collision with the port structures. Finally, the outcome of the feasibility study indicates that the proposed trajectory planner is not only theoretically sound but also practically applicable.
Future work includes an in-depth analysis of the feasibility study to refine the proposed trajectory planner, thereby enhancing its robustness and adaptability in practical applications.

% Uncomment and use as the case may be
%\begin{theorem} 
%\end{theorem}

% Uncomment and use as the case may be
%\begin{lemma} 
%\end{lemma}
 
%\appendix

\section*{Declaration of Competing Interests}
The authors declare that they have no known competing financial interests or 
personal relationships that could have appeared to influence the work reported in 
this paper.

\section*{Acknowledgment}\label{Acknowledgment}  
\normalsize{This work was supported by a Grant-in-Aid for Scientific Research from the Japan Society for Promotion of Science (JSPS KAKENHI Grant Number 22H01701).}

\printcredits

%\section{Appendix}\label{Sec:6-Appendix}   
    %\setcounter{figure}{0} 
   % \setcounter{table}{0} 
   % \renewcommand{\thetable}{\thesection.\arabic{table}}
    %\renewcommand{\thefigure}{\thesection.\arabic{figure}}

\bibliographystyle{cas-model2-names}

\bibliography{bibsource}



\end{document}